\documentstyle[12pt]{article}
\include{epsf}

\newlength{\extralineskip}

\newcommand{\beq}{\begin{equation}}
\newcommand{\eeq}{\end{equation}}
\newcommand{\bd}{\begin{displaymath}}
\newcommand{\ed}{\end{displaymath}}
\def\bea{\begin{eqnarray}}
\def\eea{\end{eqnarray}}

\def\ba{\beq\new\begin{array}{c}}
\def\ea{\end{array}\eeq}

\def\inbar{\,\vrule height1.5ex width.4pt depth0pt}
\def\IC{\relax\hbox{$\inbar\kern-.3em{\rm C}$}}
\def\IR{\relax{\rm I\kern-.18em R}}
\def\IN{\relax{\rm I\kern-.18em N}}

\def\Tr{{\rm Tr}}

\def\e{~{\rm e}}
\makeatletter
\newdimen\normalarrayskip              % skip between lines
\newdimen\minarrayskip                 % minimal skip between lines
\normalarrayskip\baselineskip
\minarrayskip\jot
\newif\ifold             \oldtrue            \def\new{\oldfalse}
\def\arraymode{\ifold\relax\else\displaystyle\fi} % mode of array entries
\def\@arrayskip{\ifold\baselineskip\z@\lineskip\z@
     \else
     \baselineskip\minarrayskip\lineskip2\minarrayskip\fi}
\def\@arrayclassz{\ifcase \@lastchclass \@acolampacol \or
\@ampacol \or \or \or \@addamp \or
   \@acolampacol \or \@firstampfalse \@acol \fi
\edef\@preamble{\@preamble
  \ifcase \@chnum
     \hfil$\relax\arraymode\@sharp$\hfil
     \or $\relax\arraymode\@sharp$\hfil
     \or \hfil$\relax\arraymode\@sharp$\fi}}
\def\@array[#1]#2{\setbox\@arstrutbox=\hbox{\vrule
     height\arraystretch \ht\strutbox
     depth\arraystretch \dp\strutbox
     width\z@}\@mkpream{#2}\edef\@preamble{\halign \noexpand\@halignto
\bgroup \tabskip\z@ \@arstrut \@preamble \tabskip\z@ \cr}%
\let\@startpbox\@@startpbox \let\@endpbox\@@endpbox
  \if #1t\vtop \else \if#1b\vbox \else \vcenter \fi\fi
  \bgroup \let\par\relax
  \let\@sharp##\let\protect\relax
  \@arrayskip\@preamble}

\begin{document}
\thispagestyle{empty}
\rightline{\baselineskip=12pt\vbox{\halign{&#\hfil\cr &
hep-th/9612030 
&\cr %{   }&\cr
& December 1996 {   }&\cr }}}
\vskip2cm
\begin{center}
{\bf DECONFINEMENT TRANSITION FOR QUARKS ON A LINE}\\
\vskip1cm
{\bf C.R. Gattringer, L.D. Paniak and G.W. Semenoff}\\
\bigskip

{\it Department of Physics and Astronomy,\\
University of British Columbia\\6224 Agricultural Road\\
Vancouver, British Columbia, Canada V6T 1Z1}
\vskip4cm
\begin{abstract}
We examine the statistical mechanics of a 1-dimensional gas of both
adjoint and fundamental representation quarks which
interact with each other through 1+1-dimensional $U(N)$ gauge fields. 
Using large-$N$ expansion we
show that, when the density of fundamental quarks is small, there is a
first order phase transition at a critical temperature and adjoint
quark density which can be interpreted as deconfinement.  When the
fundamental quark density is comparable to the adjoint quark density,
the phase transition becomes a third order one. We formulate a way to
distinguish the phases by considering the expectation values of high 
winding number Polyakov loop operators.
\end{abstract}
\end{center}
\newpage
\setcounter{page}1
\section{Introduction}

The classical Coulomb gas is an important model in statistical
mechanics.  It is exactly solvable in one dimension. In two dimensions
it exhibits the Berezinsky-Kosterlitz-Thouless phase transition which
is the prototype of all phase transitions in two-dimensional systems
which have $U(1)$ symmetry.  In this paper, we shall discuss a
generalization of the classical Coulomb gas to a system of quarks
which interact with each other through non-Abelian electric fields.
This model is known to be exactly solvable in some special cases, for
$SU(2)$ gauge group and fundamental representation quarks in one
dimension ~\cite{nambu} and for $SU(N)$ gauge group in the large-$N$
limit with adjoint representation quarks in one dimension
~\cite{stz,sz}.  It can also be formulated on the lattice and solved
with adjoint quarks in the large $N$ limit in higher dimensions
~\cite{sz,Zar} where it has a substantially more complicated
structure, although, even there, a solution of some special cases of
the model are relevant to the deconfinement transition of three and
four-dimensional Yang-Mills theory ~\cite{dadda}.

In the present paper, we shall concentrate on solving a more general
version of the one-dimensional model than has previously been
considered and elaborating on the properties of the solution.  The 
one-dimensional case has the advantage of being directly related to a
continuum field theory, the heavy quark limit of 1+1-dimensional
quantum chromodynamics (QCD).  Part of the motivation for this work is
to study the possibility of a confinement-deconfinement phase
transition at high temperature or density in a field theory which has
some of the features of QCD, i.e.  with similar gauge symmetries and
where interactions are mediated by non-Abelian gauge fields.  QCD
exhibits confinement at low temperature and density with elementary
excitations being color neutral particles - mesons and baryons.  On
the other hand, at high temperature or high density it is very
plausible that the dynamical degrees of freedom would be quarks and
gluons - which would form a quark-gluon plasma, rather than the mesons
and baryons of low temperature nuclear physics.  At some intermediate
temperature or density there should be a crossover between these two
regimes.  There are few explicitly solvable models where this behavior
can be studied directly.  Previous to the model of ~\cite{stz,sz},
known explicit examples of phase transitions in Yang-Mills theory were
not associated with confinement, but were either lattice artifacts
\cite{gw}  unrelated to the continuum gauge theory or were
associated with topological degrees of freedom in Yang-Mills theory on
the sphere ~\cite{dougkaz} and cylinder ~\cite{dadda,matgro} and
demark a finite range of coupling constants within which the gauge
theory resembles a string theory.

In finite temperature Yang-Mills theory (or QCD with only adjoint
quarks), confinement is thought to be governed by the realization of a
global symmetry which is related to the center of the gauge group and
implemented by certain topologically non-trivial gauge transformations
that appear only at finite temperature \cite{pol,sus,sy}.  The
Polyakov loop operator is an order parameter for spontaneous breaking
of this center symmetry and yields a mathematical way of
distinguishing the confining and deconfined phases.  When fundamental
representation quarks are present, the center symmetry is broken
explicitly and the Polyakov loop operator is no longer a good order
parameter for confinement.  Whether, in this case, a mathematical
distinction of confined and deconfined phases exists, and indeed
whether there is a distinct phase transition at all, is an open
question.

Here, we will consider a toy model which resembles two-dimensional QCD
with heavy adjoint and fundamental representation quarks.  It could
also be thought of as the heavy quark limit of dimensionally reduced
higher dimensional QCD where the adjoint particles are the gluons of
the compactified dimension which get a mass 
(similar to a Debye mass) from the dimensional reduction, 
and the fundamental representation
particles are the quarks.  We will solve this model explicitly in the
large-$N$ limit.  The model with only adjoint quarks was solved in
refs.~\cite{stz,sz} and it was found that the explicit solution has a
first order phase transition between confining and deconfining phases.
These phases could be distinguished by the expectation value of the
Polyakov loop operator, which provided an order parameter for
confinement in that case.  In this paper, we shall add fundamental
representation quarks. Then, as in QCD, the center symmetry is
explicitly broken and the Polyakov loop is always non-vanishing.  We
nevertheless find that the first order phase transition persists when
the density of fundamental quarks is sufficiently small.  When the
density of fundamental quarks is increased until it is comparable to
the density of adjoint quarks, the phase transition becomes a second
order one.  When the fundamental quark density is increased further,
the phase transition is third order.

It is tempting to identify this third order phase transition as the
vestige of the deconfinement transition.  Indeed, we shall present
arguments for this.  We shall find a mathematical way to distinguish
the phases on either side of the third order transition based on the
behavior of Polyakov loops with high winding numbers.  We conjecture
that the generalization of this argument to higher dimensional systems
would provide a mathematical characterization of confinement in finite
temperature QCD with fundamental representation quarks.

Another motivation of the present paper, as well as Ref.~\cite{stz,sz}
is to study a suggestion by Dalley and Klebanov \cite{dk} and
Kutasov \cite{kutasov} that 
1+1-dimensional adjoint QCD would be the simplest gauge theory model which
exhibits some of the stringy features of a confining gauge theory.  It
is a long-standing conjecture that the confining phase of a gauge
theory can be described by a string theory \cite{polyakov1}.  There
are only two cases where this relationship is well understood, 
two-dimensional Yang-Mills theory ~\cite{kk,kkk,gross} and compact quantum
electrodynamics ~\cite{polyakov2}.  At low temperatures 
1+1-dimensional adjoint QCD is confining in the conventional sense that
quarks only appear in the spectrum in color neutral bound states.
This is a result of the fact that, in one dimension, the gluon field
has no propagating degrees of freedom and therefore it cannot form a
color singlet bound state with an adjoint quark.  As a result, the
quantum states are color singlet bound states of two or more adjoint
quarks.  The spectrum contains an infinite number of families of
multi-quark bound states which resemble asymptotically linear Regge
trajectories ~\cite{kutasov,dk,Bhanot} and, for large energies, the
density of states increases exponentially with energy ~\cite{kogan}.
This implies a Hagedorn transition ~\cite{hagedorn} at high
temperature.  Kutasov ~\cite{kutasov} supported this view by using an
argument originally due to Polchinski ~\cite{polchinski} that a
deconfinement transition occurs when certain winding modes become
tachyonic at high temperature.  This behaviour was a feature of the
explicit first order deconfinement transition found in the model
considered in ~\cite{stz,sz}.  That model, which coincides with the
heavy quark limit of adjoint QCD, is effectively a statistical
mechanical model for strings of electric flux, with quarks attached
to their ends.

The large-$N$ expansion of two-dimensional adjoint QCD has the same
complexity as the large-$N$ expansion of a higher dimensional
Yang-Mills theory and the leading order, infinite-$N$ limit cannot be
found analytically~\cite{thooft}.  In fact, the dimensional reduction
of three or four-dimensional Yang Mills theory produces 
two-dimensional QCD with massless adjoint scalar quarks, so the
combinatorics of planar diagrams is very similar.

\subsection{Overview}

This paper is organized as follows: In Section 2.1 we identify the
gauged principal chiral model which corresponds to the gas with 
sources in various representations. In Section
2.2 the quantum mechanical formulation of this unitary matrix
model is analyzed. This is followed by a section (2.3) where we rewrite the
model in terms of collective variables (eigenvalue density), which are
convenient for analyzing the large-$N$ limit.

Parametrized solutions to the collective field equations are given in 
Section 3.1,
and the parametrized free energy and its derivatives are obtained in 3.2.
We also show that they give rise to a first order differential equation
the free energy has to obey. In Section 3.3 we establish the existence
of a third order line in the phase diagram, and compute the point where
it terminates. Using numerical techniques we show in 3.4 that the critical
line continues from that point as a first order line.

In Section 4.1 we derive the contributions of the gauge field and
the sources respectively to the energy density. This is followed by
a section where we establish that higher windings of Polyakov loop 
operators behave differently in the two phases (4.2). This allows to
discuss the phase diagram using energy densities and higher winding
loop operators (4.3). The paper ends with a summary and outlook
(4.4). Some of 
the more technical calculations are given in an appendix.

\section{Formalism}
\setcounter{equation}{0}
\subsection{Effective action}
The partition function of 1+1-dimensional Yang-Mills theory at
temperature $T$ and coupled to a number $K$ of non-dynamical quarks at
positions $x_1\ldots,x_K$ in representations $R_1,\ldots,R_K$ of the
gauge group is obtained by taking the thermal average of an ensemble of
Polyakov loop operators
\begin{equation}
Z[T;x_1,\ldots,x_K;R_1,\ldots,R_K]~=~\int dA_\mu
~e^{-S[A]}~\prod_{i=1}^K{\rm Tr} {\cal P}e^{i
\int_0^{1/T}d\tau A^{R_i}_0(\tau,x_i)} \; ,
\label{ftpi}
\end{equation}
where the Euclidean action is
\begin{equation}
S[A]~=~\int_0^{1/T}d\tau
\int dx\frac{1}{2e^2}{\rm
Tr}\left(F_{\mu\nu}(\tau,x)\right)^2 \; ,
\label{Seucl}
\end{equation}
the gauge fields are Hermitean $N\times N$ matrix valued vector fields
which have periodic boundary conditions in imaginary time
\[
A_\mu(\tau,x)=A_\mu(\tau+1/T,x) \; , 
\nonumber
\]
and the field strength is
\[
F_{\mu\nu} \; \equiv \; \partial_\mu A_\nu-\partial_\nu A_\mu-
i\left[ A_\mu,A_\nu \right] \; .
\nonumber
\]
The gauge field can be expanded in basis elements of the Lie algebra $A^{R_i}_\mu \equiv A^a_\mu T^a_{R_i}$ with $T^a_{R_i}$ the generators
in the representation $R_i$. For concreteness, we consider $U(N)$
gauge theory and denote the generators in the fundamental representation as
$T^a$ with $a=1,\ldots,N^2$. They obey
\begin{equation}
\left[ T^a, T^b \right] \; = \; if^{abc}T^c \; ,
\label{gen1}
\end{equation}
normalized so that
\begin{equation}
{\rm Tr} \; T^a T^b~=~\frac{1}{2}\delta^{ab} \; ,
\label{gen2}
\end{equation}
and with the sum rule
\begin{equation}
\sum_{a=1}^{N^2} T^a_{ij}T^a_{kl}~=~\frac{1}{2}\delta_{jk}\delta_{il} \; .
\label{gen3}
\end{equation}
We remark that group elements $g^{Ad}$ in the adjoint representation
are related to the fundamental representation matrices $g$ by
\begin{equation}
(g^{Ad})^{a b} \; = \; 2 \; \mbox{Tr} ( g^\dagger T^a g T^b ) \; .
\label{gadj}
\end{equation}

The expression (\ref{ftpi}) can be obtained by canonical quantization
of 1+1-dimensional Yang-Mills theory with Minkowski space action
coupled to some non-dynamical sources
\begin{equation}
S=-\int dtdx~\frac{1}{2e^2}{\rm Tr}F_{\mu\nu}F^{\mu\nu}~+~{\rm source~
terms} \; .
\label{sourceaction}
\end{equation}
In the following, we will review an argument for representing the
partition function of Yang-Mills theory as a gauged principal chiral
model which was first given in ~\cite{gsst} and which was generalized
to the case of Yang-Mills theory with sources in ~\cite{stz,sz}.  As
is usual in canonical quantization of a gauge theory, the canonical 
conjugate $E(x)$ of the spatial component of the gauge field 
(which we denote by $A(x)$), is proportional to the electric field,
 $$E \; = \; \frac{1}{e^2}F_{01} \; ,$$ and obeys the commutation relation \begin{equation}
\left[ A^a(x), E^b(y) \right]~=~i\delta^{ab}\delta(x-y) \; .
\label{cancom}
\end{equation}
The Hamiltonian is
\begin{equation}
H=\int dx~ \frac{e^2}{2} \sum_{a=1}^{N^2} (E^a(x))^2 \; .
\label{ham}
\end{equation}
This Hamiltonian must be supplemented by the Gauss' law constraint
equation which is the equation of motion for $A_0$ following from
(\ref{sourceaction}) and which contains the color charge densities of
the sources
\begin{equation}
{\cal G}^a(x)\equiv \left( \frac{d}{dx}E^a(x) -
f^{abc}A^b(x)E^c(x) + \sum_{i=1}^K T^a_{R_i}\delta(x-x_i)\right)
\sim 0 \ .
\label{gauss}
\end{equation}
Here, the particles with color charges are located at positions
$x_1,\ldots ,x_K$.  $T^a_{R_i}$ are generators in the representation
$R_i$ operating on the color degrees of freedom of the {\it i}'th
particle.

There are two options for imposing this constraint. The first is to
impose another gauge fixing condition such as $$A ~\sim~0\: ,$$ and to
use the constraints to eliminate both $E$ and $A$.  The resulting
Hamiltonian is
\begin{equation}
H=\sum_{i<j,a}\frac{e^2N}{4}T^a_{R_i}\otimes T^a_{R_j}
\left| x_i-x_j\right| \; ,
\label{nambuham}
\end{equation}
which was considered in Ref.~\cite{nambu}.  It is the energy of an
infinite range spin model where the spins take values in the Lie
algebra of $U(N)$.

The other option, which makes the closest contact with string dynamics,
is to impose the constraint (\ref{gauss}) as a physical state
condition,
\[
{\cal G}^a(x)~\Psi_{\rm phys}~=~0 \; .
\]
To do this, it is most illuminating to work in the functional
Schr\"odinger picture, where the states are functionals of the gauge
field, $\psi[A]$ and the electric field is the functional derivative
operator $$ E^a(x)\Psi[A]~=~\frac{1}{i}\frac{\delta}{\delta
A^a(x)}~\Psi[A] \; , $$ The time-independent functional Schr\"odinger
equation is
\[
\int dx\left( -\frac{e^2}{2}\sum_{a=1}^{N^2}
\frac{\delta^2}{(\delta A^a(x))^2}\right)
~\Psi^{a_1\ldots a_K}\left[A;x_1, \ldots, x_K\right]~
  =~{\cal E}~\Psi^{a_1\ldots
a_K}\left[A;x_1, \ldots, x_K\right] \; .
\]
Gauss' law implies that the physical states, i.e. those which obey the
gauge constraint (\ref{gauss}), transform as
\[
\Psi^{a_1\ldots a_K}\left[ A^g;x_1, \ldots, x_K\right] \; \; = \; \;
g^{\rm R_1}_{a_1b_1}(x_1)\ldots g^{\rm R_K}_{a_Kb_K}(x_K)
\; \Psi^{b_1\ldots b_K}\left[A;x_1, \ldots, x_K\right] \; ,
\]
where $$ A^g \; \; \equiv \; \; 
gAg^{\dagger}-ig\nabla g^{\dagger} \; , $$ is the gauge transform of $A$.

For a fixed number of particles, the quantum mechanical problem is
exactly solvable.  For example, the wavefunction of a fundamental
representation quark-antiquark pair is
\begin{equation}
\Psi^{ij}[A;x_1,x_2] \; = \; \left( {\cal P}e^{i\int_{x_1}^{x_2}dyA(y)}
\right)^{ij} \; ,
\label{wf1}
\end{equation}
where the path ordered phase operator represents a string of electric
flux connecting the positions of the quark and anti-quark.  The energy
is $\frac{e^2N}{4}\vert x_1-x_2\vert$.  For a pair of adjoint
quarks, the wavefunction is
\begin{equation}
\Psi^{ab}[A;x_1,x_2] \; = \; {\rm Tr}\left( T^a{\cal P}e^{i\int_{x_1}^{x_2}A}
T^b{\cal P}e^{i\int_{x_2}^{x_1}A} \right) \; .
\label{wf2}
\end{equation}
with energy $\frac{e^2N}{2} \vert x_2-x_1\vert$.  These energy states
are identical to what would be obtained by diagonalizing the `spin'
operators in the gauge fixed Hamiltonian (\ref{nambuham}).

Note that the wavefunctions (\ref{wf1}) and (\ref{wf2}) are not
normalizable by functional integration over $A$.  This is a result of
the fact that the gauge freedom has not been entirely fixed, so that
the normalization integral still contains the infinite factor of the
volume of the group of static gauge transformations.

In general, for a fixed distribution of quarks, a state-vector is
constructed by connecting them with appropriate numbers of strings of
electric flux so that the state is gauge invariant.  The number of
ways of doing this fixes the dimension of the quantum Hilbert space.
If the flux strings overlap, the Hamiltonian can mix different
configurations, so the energy eigenstates are superpositions of string
configurations.  However, this mixing is suppressed in the large-$N$
limit (i.e. the strings are non-interacting) and any string
distribution is an eigenstate of the Hamiltonian with eigenvalue
$(e^2N/4)\times$(total length of all strings).

We shall study the thermodynamics of this system by constructing the
partition function.  We work with the grand canonical ensemble and
assume that the quarks obey Maxwell-Boltzman statistics.  The
partition function of a fixed number of quarks is constructed by
taking the trace of the Gibbs density $e^{-H/T}$ over physical states.
This can be implemented by considering set of all states in the
representation of the commutator (\ref{cancom}), spanned by, for
example, the eigenstates of $A^a(x)$ and an appropriate basis for the
quarks $$\vert A \rangle\otimes e_{a_1}\otimes
e_{a_2}\otimes\ldots\otimes e_{a_K}~~.$$ Projection onto physical,
gauge invariant states is done by gauge transforming the state at one
side of the trace and then integrating over all gauge transformations
(and then dividing by the infinite volume of the gauge group)
~\cite{gpy}. The resulting partition function is
\begin{equation}
Z[T/e^2;x_1,\ldots,x_K] \; = \; \int[dA][dg]~\left<A\right| e^{-H/T}
\left|A^g\right> {\rm Tr}~g^{ R_1}(x_1)\ldots {\rm Tr}~g^{ R_K}(x_K)
~~,
\label{hkpi}
\end{equation}
where $[dg]$ is the Haar measure on the space of mappings from the
line to the group manifold and $[dA]$ is a measure on the convex
Euclidean space of gauge field configurations.  The expression
(\ref{hkpi}) is identical to (\ref{ftpi}) with the Polyakov loop
operator is the trace of the group element $g(x)$ in the appropriate
representation.

In going over to the grand canonical ensemble the first step is to
integrate over all particle positions. We then multiply by the
fugacities for each type of charge: a factor of $\lambda_R$ for each
quark in representation $R$.  To impose Maxwell-Boltzmann statistics,
we divide by the factorial of the number of quarks in each
representation.  We then sum over all numbers of quarks in each
representation.  This exponentiates the fugacities, resulting in the
grand partition function
\begin{equation}
Z[T/e^2,\lambda_R]~=~\int [dA][dg]~e^{-S_{\rm eff}[A,g]} \; ,
\label{partition}
\end{equation}
where the effective action is
\begin{equation}
\exp \Big( -S_{\rm eff}[A,g] \Big) = \left< A\right| e^{-H/T}\left| A^g
\right>\exp \left(\int dx~\sum_{R}\lambda_R {\rm Tr}~ g^R(x) \right) \; ,
\label{Seff0}
\end{equation}
and the summation in the exponent is over all the irreducible
representations of $U(N)$ we want to consider.  
The Hamiltonian is the Laplacian on the
space of gauge fields.  The heat kernel obeys the equation
\[
\left( T^2\frac{\partial}{\partial T}+\int dx\frac{e^2}{2}
\sum_{a=1}^{N^2} \left( \frac{\delta}{\delta A^a(x)} \right)^2
\right)\left< A\right|
e^{-H/T}\left| A^g\right> \; = \; 0 \; ,
\]
with the boundary condition
\[
\lim_{1/T\rightarrow0}\left< A\right| e^{-H/T}\left|A^g\right>
\; \; = \; \; \delta(A-A^g) \; .
\]
These equations are easily solved by a Gaussian - divided by a
$T$-dependent constant: $$
\left< A\right| e^{-H/T} \left|
A^g\right> \; \; \sim \; \;
\exp\left(-\int dx ~\frac{T}{e^2}~{\rm Tr}~(A-A^g)^2\right)\ .
$$

We see that the effective theory is the gauged principal chiral model
with a potential energy term for the group-valued degrees of freedom,
\begin{equation}
S_{\rm eff}[A,g]~=~ \int dx~\left( \frac{N}{2 \gamma}~ {\rm Tr}\Big|
\nabla g(x) + i[A(x),g(x)]\Big|^2 -\sum_{R}\left(
\lambda_R
{\rm Tr}~g^R(x)  \right)\right) \label{seff} \; .
\label{seffrep}
\end{equation}
Note that we have introduced the coupling constant
\begin{equation}
\gamma \; \; \equiv \; \; \frac{ e^2 \; N}{2 T} \; .
\label{gammadefn}
\end{equation}
When we analyze the limit $N \rightarrow\infty$, we will tune $e^2$
such that $\gamma$ is constant. Moreover, we assume that 
the fugacities $\lambda_R$
are scaled such that all terms in the action (\ref{seffrep}) are of order
$N^2$.

The potential energy term in the effective action,
\begin{equation}
V(g)~\equiv~-\sum_R \lambda_R {\rm Tr}\left(g^R(x)\right) \; ,
\label{poten}
\end{equation}
is the expansion of a local class function of the group element $g(x)$
(one which obeys $V(g)=V(hgh^{-1})$ for $h\in U(N)$) in group
characters with coefficients $\lambda_R$.  The characters
\[
\chi_R(g)~\equiv~{\rm Tr}\left( g^R(x) \right) \; ,
\]
form a complete set of orthonormal class functions of the group
variable, with inner product
\[
\int [dg]\chi_R^*(g)\chi_{R'}(g)~=~\delta_{R,R'} \; .
\]
Here [dg] is not a functional integral measure, but is the Haar
measure for integration on $U(N)$. From the potential, we can find a
fugacity by
\[
\lambda_R~=~-\int[dg]\chi_R^*(g)V(g) \; .
\]
By tuning the fugacities appropriately, we could obtain any local
invariant potential.

The effective action (\ref{seff}) with all $\lambda_R=0$ was discussed
by Grignani et.al.~\cite{gsst} and was solved explicitly in the limit
$N\rightarrow \infty$ by Zarembo ~\cite{Zar}. The model with
$\lambda_{Ad}\neq 0$ (with adjoint quarks) was solved in
Refs.\cite{stz} and ~\cite{sz}.  The effective action (\ref{seff}) is
gauge invariant, $$S_{\rm eff}[A,g]~=~ S_{\rm eff}[A^h,
hgh^{\dagger}]~~. $$
It is also covariant under the global transformation
\begin{equation}
S_{\rm eff}[A,zg,\lambda_R]~=~ S_{\rm eff}[A,g,z^{-C_1(R)}\lambda_R] \; .
\label{zsym}
\end{equation}
where $z$ is a constant element from the center of the gauge group,
which for $U(N)$ is $\sim$ $U(1)$ and would be the discrete group
$Z_N$ for gauge group $SU(N)$. Here, $C_1(R)$ is the linear Casimir
invariant of the representation $R$, which is the number of boxes in
the Young tableau corresponding to $R$.  When the gauge group is $SU(N)$
and the only non-zero fugacities are for the zero `N-ality'
representations, i.e. those for which $C_1(R)=0$ mod $N$, there is a
global $Z_N$ symmetry.  For gauge group $U(N)$, this occurs only when
all representations with non-zero fugacities have equal numbers of
quarks and anti-quarks.  The fugacities of other non-symmetric
charges, can be thought of as an external field which breaks the
center symmetry of the system explicitly. This situation is akin to
the effect of an external magnetic field on a spin system.

\subsection{Matrix quantum mechanics}

If we re-interpret $x$ as Euclidean time, the partition function that
we have derived has the form of a Euclidean space representation of
the partition function for matrix quantum mechanics, where the free
energy is identical to the ground state energy of the matrix quantum
mechanics.  We can study the latter model by mapping the problem to
real time $\tau$ by setting $x = i\tau$ and $A\rightarrow -iA$.  The
action in real time is then
\[
S_{QM} \; = \;
\int d\tau \left( \frac{N}{2\gamma}{\rm Tr}\left|\dot g+i\left[
A,g\right]\right|^2 -V(g)\right) \; .
\]
We remark that this action must not be confused with the action
(\ref{Seucl}). $S_{QM}$ is the action for a 0+1-dimensional problem
(quantum mechanics), while (\ref{Seucl}) is the action for Yang Mills
theory in 1+1 dimensions. This remark also holds for the Hamiltonian
below. In order to avoid confusion, we label the quantum mechanical
quantities with the subscript $QM$.

The canonical momentum conjugate to the group valued position variable
$g$ is the Lie algebra element
\[
\Pi \; = \; 
\frac{N}{\gamma}\left( ig^{\dagger}\dot g +g^{\dagger}Ag-A\right) \; ,
\]
and the Hamiltonian is
\begin{equation}
H_{QM} \; = \; \frac{\gamma}{2N}{\rm Tr}\Pi^2-\frac{\gamma}{N}{\rm
Tr}\Pi(g^{\dagger}Ag-A) ~+~V(g) \; .
\label{Cqmham}
\end{equation}
The gauge field $A$ plays the role of a Lagrange multiplier which
enforces the constraint
\[
g\Pi g^{\dagger}-\Pi\sim0 \; ,
\]
and the Hamiltonian reduces to
\[
H_{QM} \;  = \; \frac{\gamma}{2N}{\rm Tr}\Pi^2 +V(g) \; .
\]
We can expand the canonical momentum as
\[
\Pi~=~ \sum_a \Pi^a T^a \; ,
\]
Then, the components satisfy the Lie algebra
\begin{eqnarray}
\left[ \Pi^a, \Pi^b\right] \; &=& \; if^{abc}\Pi^c \; , \\
\left[ \Pi^a,g \right] \; &=& \; gT^a \; , \\
\left[ \Pi^a, g^{\dagger} \right] \; &=& \;  -T^a g^{\dagger} \; .
\end{eqnarray}
It follows that in the Schr\"odinger picture the components of
the canonical momentum are represented as
\[
\Pi^a~=~{\rm Tr}gT^a\frac{\partial}{\partial g}~=~g_{ij}T^a_{jk}
\frac{\partial}{\partial g_{ik}} \; .
\]
Denoted in components the constraint reads
\[
G^a \; \equiv \; \left( {\rm Tr}T^a gT^b g^{\dagger}-
\frac{1}{2}\delta^{ab} \right)
\Pi^b \sim0 \; .
\]
The constraint has no operator ordering ambiguity.  It generates the
adjoint action of the symmetry group
\[
\left[ G^a,g\right] \; = \; \frac{1}{2} \left[ T^a, g\right] \; .
\]
The constraint can be realized as a physical state condition
\[
G^a~\psi_{\rm phys}~=~0 \; .
\]
In the representation where states are functions of $g$, this implies
that the physical states are class functions
\[
\psi_{\rm phys}(g)~=~\psi_{\rm phys}(hgh^{-1}) \; ,
\]
where $h\in U(N)$. This means that the physical states are functions
of the eigenvalues of $g$.  In a basis where $g$ is diagonal,
\begin{equation}
g~=~\left( \matrix{ e^{i\alpha_1} &0& 0 & 0&\ldots &0\cr 0
&e^{i\alpha_2}&0&0&\ldots&0\cr 0&0& e^{i\alpha_3}&0&\ldots&0\cr
0&0&\ldots & & & 0\cr 0&0&\ldots & & & e^{i\alpha_N}\cr }\right) \; ,
\label{diagg}
\end{equation}
the wavefunctions are functions of $\alpha_i$,
\begin{equation}
\psi_{\rm phys}(\alpha_1,\ldots,\alpha_N)=\psi_{\rm phys}(\alpha_1,
\ldots,\alpha_i+2\pi,\alpha_N) \; .
\label{psialph}
\end{equation}
Denoting the gauge group Laplacian in components
\begin{equation}
\triangle \; \; \equiv \; \; \sum_{a=1}^{N^2} ( \Pi^a )^2 \; ,
\label{glap}
\end{equation}
the Hamiltonian reads
\[
H_{QM} \; \; = \; \; \frac{\gamma}{4N}\Delta \; + \; V(g) \; .
\]
Since the potential $V(g)$ is also a class function and depends only
on the eigenvalues, when operating on the physical states, the
Hamiltionian can be expressed in terms of eigenvalues and derivatives
by eigenvalues
\[
H_{QM} \; =
\; \frac{\gamma}{4N}\frac{1}{\tilde J(\alpha)}\left(\sum_1^N
-\frac{\partial^2} {\partial\alpha_i^2}-N(N^2-1)/12\right)
\tilde J(\alpha)+V(\alpha) \; ,
\]
where
\[
\tilde J(\alpha)=\prod_{i<j} 2\sin\frac{1}{2}(\alpha_i-\alpha_j)
=\frac{1}{(2i)^{N(N-1)/2}}\frac{ J(\alpha)}{\prod_i z_i^{(N-1)/2} } \; ,
\]
and
\[
J(z)=\prod_{i<j}(z_i-z_j)~~~,~~z_i\equiv e^{i\alpha_i} \; .
\]

The physical states must by symmetric functions of $\alpha_i$. (There
is a residual gauge invariance \cite{lang} under the Weyl group which
permutes the eigenvalues and the physical state condition requires
that the physical states be symmetric under these permutations.)
The normalization integral for the wavefunction is
\[
\int [dg] \psi^{\dagger}(g)\psi(g)~=~1 \; .
\]
Since the integrand depends only on the eigenvalues of $g$, It is
convenient to write the Haar measure as an integral over eigenvalues
of $g$ with the Jacobian which is the Vandermonde determinant,
\[
\int (\prod_i d\alpha_i) \vert \tilde J(\alpha)\vert^2
\psi^{\dagger}(\alpha)
\psi(\alpha)~=~1 \; .
\]
The Hamiltonian and inner product have a particularly simple form when
we redefine the wavefunction as
\[
\tilde\psi(\alpha_1,\ldots,\alpha_N)\equiv \tilde J(\alpha) \psi(\alpha_1,
\ldots,\alpha_N) \; .
\]
Since $\tilde J$ is antisymmetric, $\tilde\psi$ is a completely
antisymmetric function of the eigenvalues, which we can think of as
the coordinates of fermions.  The Hamiltonian is that of an
interacting Fermi gas
\[
\left\{
\frac{\gamma}{4N}\left( \sum_1^N -\frac{\partial^2}{\partial\alpha_i^2}
-N(N^2-1)/12\right)+V(\alpha)\right\}\tilde\psi(\alpha)={\cal E}\tilde
\psi(\alpha) \; .
\]
This correspondence of a c=1 matrix model with a Fermi gas was first
pointed out in Ref.~\cite{bipz}.
\subsection{Large N: Collective variables}

In this section we shall examine the collective field formulation of the
large-$N$ limit of the theory that we discussed in the last subsection
\cite{Zar,JS}. The Hamiltonian obtained in the last subsection reads
\begin{equation}
H_{QM}  \; \; = \; \; \frac{\gamma}{4 N} 
\sum_{a=1}^{N^2} ( \Pi^a )^2
 \; + \; V(g) \; ,
\label{ham2}
\end{equation}
with (compare (\ref{poten}))
\[
V(g)~\equiv~-\sum_R \lambda_R {\rm Tr}\left(g^R(\tau)\right) \; .
\]
It was shown (compare (\ref{diagg}),(\ref{psialph}))
that the wavefunction depends only on the eigenvalues $e^{i\alpha_j}$
of $g$ and thus the density of eigenvalues
\[
\rho(\theta,\tau)~\equiv~\frac{1}{N}\sum_{i=1}^N
\delta(\theta-\alpha_i(\tau)) \; ,
\]
completely characterizes the properties of the system.
Interpretation of the physics of the system at large $N$
is more convenient
when one considers the Fourier transform of the eigenvalue distribution
\begin{equation}
\rho(\theta,\tau)~=~\frac{1}{2\pi}~+~
\frac{1}{2\pi}
\sum_{n \neq 0} c_n(\tau) e^{-in\theta} \; ,
\label{den}
\end{equation}
where we have defined
\[
c_n(\tau) \; \equiv \; \frac{1}{ N} \Tr g^n(\tau) ~~~~,~~~
c_{-n}(\tau) \; = \; \overline{c_n(\tau)} \; .
\]
We now turn our attention to developing the collective field theory
formulation of the $\mbox{(thermo-)}$ dynamical problem given by the
Hamiltonian (\ref{ham2}).  Since the wavefunction depends only on
the eigenvalues of $g$, we would like a Hamiltonian
equivalent to (\ref{ham2}) but written in terms of
the eigenvalue density $\rho$ and a conjugate momentum $\Pi$.
At large $N$ we will find this Hamiltonian and write equations
of motion for $\rho$ and $\Pi$. So far we have not imposed any restriction 
on the potential $V(g)$, but from now on we assume, that it can be 
expressed as a functional of the eigenvalue density $\rho(\theta)$. 
In particular 
the potential we are going to analyze below will have this property.

The canonical momentum operates on the wavefunction as
\begin{eqnarray}
\Pi^a\psi[\rho]~&=&~\int d\theta\left[\Pi^a,\rho(\theta)\right] \frac{\delta}
{\delta\rho(\theta)}~\psi[\rho]
\nonumber\\
&=&\frac{1}{2\pi N}\int d\theta\sum_K e^{-iK\theta}K{\rm Tr}(T^ag^K)
\frac{\delta}{\delta\rho(\theta)}~\psi[\rho] \; ,
\nonumber
\end{eqnarray}
and the Laplacian (\ref{glap}) is
\begin{eqnarray}
\Delta\psi[\rho]~=~\left(
\frac{1}{4\pi N}\int d\theta\sum_K e^{-iK\theta}\left| K\right|\left(
\sum_{L=0}^K {\rm Tr}g^L {\rm Tr}g^{K-L}-N{\rm Tr}g^K\right)
\frac{\delta}{\delta\rho(\theta)}
\right.\nonumber\\ \left.
+ \; \frac{1}{8\pi^2N^2}\int d\theta d\theta'
\sum_{KL}K L e^{-iK\theta-iL\theta'} {\rm
Tr}g^{K+L}~\frac{\delta^2}{\delta\rho(\theta)\delta\rho(\theta')}
\right)~\psi[\rho] \; ,
\nonumber
\end{eqnarray}
which can be written as
\begin{eqnarray}
\Delta\psi[\rho] \; &=& \;
-\frac{1}{2N}\int d\theta\rho(\theta)\left\{\left( \frac{\partial}
{\partial\theta}\frac{\delta}{\delta\rho(\theta)}\right)^2-
N^2{\cal P}\int d\theta' \rho(\theta')\cot\left(
\frac{\theta-\theta'}{2}\right)\frac{\partial}{\partial\theta}
\frac{\delta}{\delta\rho(\theta)}
\right\}\psi[\rho]
\nonumber\\
&=& \; -\frac{1}{2N}\int d\theta\rho(\theta)\left(\left(
\frac{\partial}{\partial\theta}
\frac{\delta}{\delta\rho(\theta)}+{\cal V}(\theta)\right)^2 -
{\cal V}^2(\theta)\right)\psi[\rho] \; ,
\nonumber
\end{eqnarray}
where
\[
{\cal V}(\theta)=\frac{N^2}{2}{\cal P}\int d\theta'\rho(\theta')\cot\left(
\frac{\theta-\theta'}{2}\right) \; .
\]
${\cal P}$ indicates principal value integral.

The transformation of the wavefunction
\[
\psi[\rho] \; = \; \tilde\psi[\rho]\exp\left(-\frac{N^2}{2}\int
d\theta d\theta'
\ln\sin\frac{\left|\theta-\theta'\right|}{2} \rho(\theta) \rho(\theta')
\right) \; ,
\]
transforms the derivative in the Schr\"odinger equation so that it has
the form
\begin{equation}
\left\{-\frac{\gamma}{8N^2}\int d\theta \rho(\theta)\left\{ \left(
\frac{\partial}{\partial\theta}\frac{\delta}{\delta\rho(\theta)}
\right)^2-{\cal V}^2(\theta)\right\}+V[\rho]\right\}~\tilde\psi[\rho]
~=~E~\tilde\psi[\rho] \; .
\label{jsham}
\end{equation}
The second term on the left-hand-side of this equation has a simple
form.  In Appendix A.1 it is shown that it gives rise to a term which is cubic in the density.
Thus, the Schr\"odinger equation has the form
\begin{equation}
\left\{-\frac{\gamma}{8N^2}\int d\theta \left(\rho(\theta) \left(
\frac{\partial}{\partial\theta}\frac{\delta}{\delta\rho(\theta)}
\right)^2-N^4\frac{\pi^2}{3}\rho^3(\theta)
\right)+V[\rho]\right\}~\tilde\psi[\rho]~
=~E~\tilde\psi[\rho] \; ,
\label{jsham2}
\end{equation}
up to an overall constant.

The large-$N$ limit is dominated by the eikonal approximation. In this
approximation, we make the ansatz for the wavefunction
\[
\tilde\psi[\rho]~=~\exp\left( iN^2 S[\rho]\right) \; .
\]
The eikonal, $S$ then obeys the equation
\begin{equation}
\frac{H_{QM}[\rho,\Pi]}{N^2}~=~
\frac{\gamma}{8}\int d\theta \left\{\rho(\theta) \left(
\frac{\partial }{\partial\theta}\frac{\delta~S}{\delta\rho(\theta)}
\right)^2+\frac{\pi^2}{3}\rho^3(\theta)
\right\}+\frac{1}{N^2}V[\rho] ~=~\frac{E}{N^2} \; .
\label{schroedeqn}
\end{equation}
Here, we have ignored a term which is of subleading order in $N^2$.
We have also assumed that $V[\rho]$ will be of order $N^2$ (compare Section
2.1) and that the
natural magnitude of the energy eigenvalue is of order $N^2$.

To solve this equation for the ground state, we must find its minimum
by varying $\rho$ and the canonical momentum
$$
\Pi \; = \; \delta S/\delta\rho \; ,
$$
subject to the condition that $\rho$ is normalized.  This leads to
the equations of collective field theory
\begin{eqnarray}
\frac{\partial}{\partial \tau}
\rho(\tau,\theta) \; \; &=&  \; \;
\frac{\; \delta H_{QM}/N^2 \; }{\delta \Pi(\tau,\theta)} \; ,
\nonumber\\
\frac{\partial}{\partial \tau}
v(\tau,\theta) \; \; &=& \; \; -\frac{\partial}{\partial\theta} \;
\frac{ \; \delta H_{QM}/N^2 \; }{\delta \rho(\tau,\theta)} \; ,
\nonumber
\end{eqnarray}
where
\bd
v(\tau,\theta)~\equiv~\frac{\partial}{\partial\theta} \; 
\Pi(\tau,\theta) \; .
\ed
Taking the derivative of the second equation with respect
to $\theta$ eliminates a Lagrange
multiplier which must be introduced on order to enforce the
normalization condition for $\rho$. Using (\ref{schroedeqn}) one finds
\begin{eqnarray}
\frac{\partial}{\partial \tau}\rho \; + \; \frac{\gamma}{4}\frac{\partial}
{\partial\theta}\left(\rho v\right) \; &=& \; 0 \; ,
\nonumber\\
\frac{\partial}{\partial \tau}v \; + \; \frac{\gamma}{8}\frac{\partial}
{\partial\theta}\left( v^2+\pi^2\rho^2\right) \; + \; \frac{1}{N^2}
\frac{\partial}{\partial\theta}\frac{\delta}{\delta\rho}V[\rho] \;
&=& \; 0 \; .
\label{colfe}
\end{eqnarray}
It is interesting to note that these are nothing but Euler's
equations for a fluid with equation of state
$P= \pi^2 \rho^3 /3$ on a cylinder with coordinates
$(\theta, \tau)$. The inclusion of a potential $V(\theta,\tau)$
corresponding to non-Abelian charges is equivalent to subjecting
the fluid to an external force which is derived from $V(\theta,\tau)$.

We shall use these equations in the next section where we analyze
the large-$N$ limit of a mixed gas of adjoint and fundamental
charges.

\section{Free energy and critical behaviour}
\setcounter{equation}{0}
\subsection{Static solutions to the collective field equations}

In this section we will find static solutions to
the collective field equations (\ref{colfe}). The most simple 
potentials involve only the lowest representations, the fundamental, 
its conjugate and the adjoint. We shall consider a slight generalization 
of these and  use powers of the lowest representations to include
multiple windings of the Polyakov loop operator. Our potential 
reads\footnote{It should be remarked that parts of this section can easily 
be extended to more general potentials.}
\begin{equation}
V(g)\; \equiv \; -\sum_{n=1}^\infty \left( \kappa_n N{\rm Tr}(g^n)+
\bar\kappa_n N
{\rm Tr}((g^{\dagger})^n)+\lambda_n\vert{\rm Tr}g^n\vert^2 \right) \; ,
\label{potential}
\end{equation}
where we made use of (compare (\ref{gen3}), (\ref{gadj}))
\[ 
{\rm Tr}~(g^{Ad}(x))^n \; = \;
\left| {\rm Tr}~ g^n(x)\right|^2 \; ,
\]
to relate the trace in the adjoint representation to the trace in
the fundamental representation. The couplings for the 
fundamental representation charges (and their conjugates) were chosen to 
scale $\sim N$, to make the potential of order $N^2$.

The potential (\ref{potential}) indeed can be expressed as
a functional of the eigenvalue density (\ref{den}). 
The collective field Hamiltonian (\ref{schroedeqn}) then reads
\bea
\frac{H_{QM}}{N^2}~
&=&~\frac{\gamma}{8} \int d\theta\left[ \rho(\theta)\left(
v(\theta) \right)^2
 +\frac{\pi^2}{3}\rho^3(\theta) \right] \label{col} \\
&-& \sum_{n=1}^\infty
\left( \lambda_n\left| \int d\theta~\rho(\theta)e^{in\theta}\right|^2
+ \kappa_n \int~d \theta~ \rho(\theta) \e^{i n \theta} + \bar{\kappa}_n
\int ~d \theta ~\rho( \theta) \e^{-i n \theta} \right) - 
\frac{\gamma}{96}\; .
\nonumber
\eea
In order to maintain correspondence with the original version of the
Hamiltonian (\ref{ham2}) we subtract the constant $\gamma/96$. It sets the 
energy scale such that the free energy vanishes in the confined phase 
of the model with only adjoint charges (see below). 

The corresponding collective field equations (\ref{colfe}) read
\begin{equation}
\frac{\partial\rho}{\partial x} \; + \;
\frac{\gamma}{4}\frac{\partial}{\partial\theta}
(\rho v) \; = \; 0 \; ,
\label{coll1}
\end{equation}
\begin{equation}
\frac{\partial v}{\partial x} \; + \; \frac{ \gamma}{8} \frac{\partial v^2}
{\partial\theta}
\; - \; \frac{ \pi^2\gamma}{8} \frac{\partial\rho^2}{\partial\theta} \; +
\; \frac{\partial}{\partial \theta} \sum_n \left[
(\lambda_n c_{-n} + \kappa_n) \e^{i n \theta} +
( \lambda_n c_n + \bar{\kappa}_n ) \e^{-i n \theta} \right]~=~0 \; .
\label{coll2}
\end{equation}
where $c_n $ are the $x$-dependent Fourier coefficients of $\rho$
as introduced in (\ref{den}).  Note that we also performed the change of
variables, $\tau \rightarrow -ix$ and
$v\rightarrow iv$ in these equations in order to invert
the Wick rotation performed at the beginning of Section 2.2 prior to
canonical quantization.

We will only consider real, static solutions of the non-linear
equations (\ref{coll2}), that is, where 
$\rho(\theta,x) = \rho_0(\theta)$ and the velocity $v$
vanishes identically.  Consequently
\vskip1mm
\begin{equation}
\rho_0(\theta)=\left\{ \matrix{
2 \sqrt{\frac{2}{\gamma\pi^2} }\sqrt{ E+\sum(\lambda_n c_{-n}+\kappa_n)
e^{in\theta}+\sum(\lambda_n c_n+ \bar{\kappa}_n )e^{-in\theta} } &
\mbox{where } \rho \mbox{ is real} \cr 0 & {\rm otherwise}\cr}\right. \; .
\label{Cdens}
\end{equation}
The constant of integration $E$ has physical interpretation as the
 Fermi energy of a collection of $N$ fermions \cite{bipz}
in the potential $V[\rho]$ and is fixed by the normalization condition
\beq
1 \; = \; \int d \theta ~\rho_0(\theta) \; .
\label{normcond}
\eeq
Here it is more convenient to express the $c_n$ in terms of $\rho_0$
(compare (\ref{den}))
\begin{equation}
c_n \; = \;  \int d \theta ~\rho_0(\theta) e^{in\theta} \; .
\label{cnphi}
\end{equation}
The real support of the function $\rho_0(\theta)$ is the positive support of 
$\Lambda \equiv E+\sum(\lambda_n c_{-n}+\kappa_n)
e^{in\theta}+\sum(\lambda_n c_n+ \bar{\kappa}_n )e^{-in\theta}$.  The
zeros of $\Lambda$ define the edges of the eigenvalue distribution and
when these zeros condense one has critical behaviour in the observables
of the model as in general Hermitean and unitary matrix models.

\subsection{A differential equation for the free energy}
In this subsection we compute all first order derivatives of the
free energy and show that they obey a differential equation of the
Clairaut type.

Inserting the static solution (\ref{Cdens}) in (\ref{col}) we obtain for the
free energy
\begin{equation}
\frac{1}{N^2} \langle H_{QM} \rangle \; \; \equiv \; \;
f \; \; = \; \; \frac{1}{3} E \; - \; \frac{1}{3} \sum_{n=1}^\infty
\Big[ \lambda_n |c_n|^2 \; + \; 2(\; \kappa_n c_n \; + \;
\bar{\kappa}_n c_{-n} \; ) \Big] - \frac{\gamma}{96} \; .
\label{CfreeE}
\end{equation}
Note that $f$ is the leading coefficient ($O(N^2)$)
of the energy for the matrix quantum mechanics problem, but for the
quark gas problem plays the role of the leading coefficient of the
energy {\it density}.

Deriving the expression (\ref{CfreeE}) with respect to $\lambda_J$ for
some fixed $J$ and using derivatives of Equations (\ref{normcond})
and (\ref{cnphi}) with respect to the same parameter one obtains
\begin{equation}
\frac{d f}{d \lambda_J} \; = \; -|c_J|^2 \; .
\label{Cdfdl}
\end{equation}
We use the notation $d/d\lambda_J$ to indicate that also the
$c_n$ and $E$ which implicitly depend on $\lambda_J$ are derived
with respect to this coupling. Similarly one can show
\begin{equation}
\frac{df}{d \kappa_J} \; = \; - c_J\; ,
\label{Cdfdk}
\end{equation}
and
\begin{equation}
\frac{df}{d\gamma} \; = \; \frac{1}{3 \gamma} E \; + \;
\frac{1}{3 \gamma} \sum_{n=1}^\infty \Big[ 2 \lambda_n |c_n|^2 \; +
\; \kappa_n c_n \; + \;
\bar{\kappa}_n c_{-n} \Big] \; - \; \frac{1}{96} \; .
\label{Cdfdg}
\end{equation}
It is interesting to notice, that combining Equations (\ref{Cdfdl}) -
(\ref{Cdfdg}) gives rise to a first order differential equation
of the Clairaut type
\begin{equation}
\gamma \frac{df}{d\gamma} \; + \; \sum_{n=1}^\infty
\Big[ \lambda_n \frac{df}{d\lambda_n} +
\kappa_n \frac{df}{d \kappa_n} + \bar{\kappa}_n
\frac{df}{d \bar{\kappa}_n} \Big] \; \; = \; \; f \; .
\label{Cdiffe}
\end{equation}
This differential equation has general solutions of the form
\[
f ( \gamma, \lambda_n, \kappa_n ) \; \; = \; \;
\gamma \; F( \frac{\lambda_n}{\gamma}, \frac{\kappa_n}{\gamma} ) \; ,
\]
where $F$ is some arbitrary smooth function. This result shows that
the parameter $\gamma$ is not driving the physical properties of the model,
but rather sets the energy scale. The differential equation (\ref{Cdiffe})
gives no further restrictions on the function $F$ and another
analysis will be adopted in the next section. However, when finding the
physical interpretation of the phase diagram in Section 4.3, the differential
equation (\ref{CfreeE}) will be a valuable tool.

\subsection{Regime of the third order phase transition}

We now restrict ourselves  to the case of only one pair of non-vanishing
couplings
$\lambda_J, \kappa_J \neq 0$. Furthermore it is sufficient to consider
$\kappa_J$ real, since an eventual phase of $\kappa_J$ can always be
removed by using the covariance (\ref{zsym}) of the action and
Haar measure $[dg]$ under transformations by a constant element of $U(1)$.

In the form of (\ref{Cdens}) it is evident we need to solve
simultaneously for  the normalization
condition (\ref{normcond}) and the Fourier moment (\ref{cnphi})
in order to
have a self-consistent solution of the saddle-point equations.
We begin  by introducing an auxiliary complex
parameter
\begin{equation}
r_J~e^{iJ \beta_J}\; \equiv \; \lambda_J ~ c_{-J} + \kappa_J \; ,
\label{Cname}
\end{equation}
and rescaling the Fermi energy as
\begin{equation}
E \; \; \equiv \; \;  2 \mu \; r_J \; .
\label{CEscal}
\end{equation}
With this notation the normalization and moment equations are respectively
\begin{equation}
1 \;  = \; 2\sqrt{\frac{r_J}{\gamma}} \; I(\mu)
\; \; \; \; \; \; , \; \; \; \; \; \;
c_Je^{i J \beta} \; = \; 2\sqrt{\frac{r_J}{\gamma}} \; M(\mu) \; ,
\label{Cnorm}
\end{equation}
where we have defined the integrals
\begin{eqnarray}
I(\mu) \; &\equiv& \; \frac{2}{\pi} \int_{-\pi}^{\pi} \sqrt{\mu +
\cos(J\theta)} \;
\mbox{H}(\mu + \cos(J\theta)) \; d\theta \; ,\nonumber \\
M(\mu) \; &\equiv& \; \frac{2}{\pi} \int_{-\pi}^{\pi} \cos{J\theta}
\sqrt{\mu + \cos(J\theta)} \;
\mbox{H}(\mu + \cos(J\theta)) \; d\theta \; .
\label{Cparint}
\end{eqnarray}
Here H(..) denotes the step function. A simple transformation of the
integration variable shows that $I(\mu)$ and $M(\mu)$ are independent of
$J$. Thus $J$ enters only as the subscript of the parameters. For
notational convenience we abbreviate
\begin{equation}
\lambda_J \; \equiv \; \lambda \; \; \; \; , \; \; \; \;
\kappa_J e^{-iJ\beta_J} \; \equiv \; \kappa \; .
\label{abbr}
\end{equation}
We remark that $I(\mu), M(\mu)$,
and thus $c_J e^{iJ\beta_J}$ are real. 
Eliminating the moment $c_J$ from
(\ref{Cnorm}) by using the definition (\ref{Cname}) we obtain
\begin{equation}
\frac{{\kappa}}{\gamma} \; \; = \; \; \frac{1}{I(\mu)^2}
\left[ \frac{1}{4} - \frac{\lambda}{\gamma} I(\mu) M(\mu) \right] \; .
\label{Cnecc}
\end{equation}
This family of lines in the $\kappa, \lambda$-plane parametrized
by $\mu$
represent a necessary condition which a solution of the normalization and
moment equations (\ref{Cnorm}) must obey. From the last equation it is 
obvious that also the product $
\kappa \; = \; \kappa_J e^{-iJ\beta} $
is real (thus $e^{iJ\beta}$ is just a sign). It occurs as a natural
parameter when rewriting the free energy in terms of $I(\mu)$ and $M(\mu)$
(use (\ref{CfreeE}))
\begin{equation}
f \; = \; \frac{\gamma}{12 I(\mu)^2} \left[
2\mu - \frac{M(\mu)}{I(\mu)} \right] \; - \;
\kappa \frac{M(\mu)}{I(\mu)} \; - \frac{\gamma}{96} \; ,
\label{Cfmu}
\end{equation}
where we have eliminated $\lambda$ using the necessary
condition (\ref{Cnecc}). Also the first derivative of the
free energy with respect to $\kappa$ can be expressed conveniently in terms of $I(\mu)$ and $M(\mu)$
\begin{equation}
\frac{d  f}{d \kappa } \; = \;  -2 \frac{M(\mu)}{I(\mu)} \; .
\label{Cdfmu}
\end{equation}
Remember that we restricted ourselves to $\kappa_J$ real, and thus we
encounter a factor 2 compared to \ref{Cdfdk}), since a real $\kappa_J$
is the same for both terms $c_J$ and $c_{-J}$ in the potential
(\ref{CfreeE}).

With the parametric solution (\ref{CfreeE}) at hand we turn our attention
to establishing the critical behaviour in this model.
In the appendix it is shown that the first derivatives of $I(\mu)$ and
$M(\mu)$ have non-analytic behaviour at $\mu = 1$, hence the expression
(\ref{Cdfmu}) suggests that the vicinity of $\mu = 1$ is a
natural place to
look for non-analytic behaviour in the free energy of our model.
Using the explicit results (\ref{Cintmu1}) for $I(1), M(1)$ and
(\ref{Cnecc}) we obtain the necessary condition for the critical
($\mu = 1$) values of $\lambda$ and $\kappa$
\begin{equation}
\frac{\kappa^c}{\gamma} \; \; = \; \;  \frac{\pi^2}{512} \; -
\; \frac{1}{3}\frac{\lambda^c}{\gamma}  \; .
\label{Ccritlin}
\end{equation}
Having identified a line in the phase space where we expect critical behaviour
we will now proceed to establish the details of this critical behaviour.
Following \cite{gubser}
we begin by expanding about $\mu =1$ and the line
(\ref{Ccritlin})
\begin{equation}
\mu = 1 + \varepsilon, \varepsilon > 0 \; \; \; \; ,  \; \; \; \;
I(1 + \varepsilon ) = I_c + \delta I \; \; \; \; , \; \; \; \;
M(1 + \varepsilon ) = M_c + \delta M \; ,
\label{Cexp}
\end{equation}
($I_c \equiv I(1), M_c \equiv M(1)$) and analyze the variation of $\kappa$
around $\kappa^c$ while keeping $\lambda$ fixed
\begin{equation}
\kappa = \kappa^c + \delta \kappa
\; \; \; \; ,
\; \; \; \; \lambda = \lambda^c \; .
\label{Cexp2}
\end{equation}
Due to the nontrivial support of the integrands in (\ref{Cparint})
in principle one has to distinguish the cases $\mu > 1$ and $\mu < 1$;
(see the appendix for details). In order to keep the formulas simple,
we explicitly analyze only the case $\mu > 1$ as given by
(\ref{Cexp}), (\ref{Cexp2}). The case $\mu < 1$ can be treated along the
same lines and we denote the corresponding results in the end.

The expansion now consists of two steps. We first
expand the necessary condition
(\ref{Cnecc}) at $\mu =1$ to obtain the relation between the variation
$\delta \kappa $ and $\varepsilon$. In the second step we
expand the right hand side of (\ref{Cdfmu}) at $\mu = 1$ and use the
result of step one to express the variation of $ df/d\kappa$
in terms of $\delta \kappa $. The latter result can then be
used to analyze eventual singular behaviour of higher derivatives of the
free energy.

Expanding the necessary condition (\ref{Cnecc}) and using (\ref{Ccritlin})
we obtain for the variation of $\kappa$ to lowest order
\begin{equation}
\frac{\delta \kappa}{\gamma} \; \; \; = \; \; \; - \frac{1}{2}
\frac{\delta I}{(I_c)^3} \; - \; \frac{\lambda^c}{\gamma}
\frac{M_c}{I_c} \left[ \; \frac{\delta M}{M_c}  -
\frac{\delta I}{I_c} \right] \; \; \; = \; \; \;
\left[ \frac{- \pi^2}{256} + \frac{4}{3} \frac{\lambda^c}{\gamma} \right]
\frac{\delta I}{I_c} \; \; - \; \;
\frac{\lambda^c}{\gamma} \frac{\varepsilon}{2} \; ,
\label{Cvar1}
\end{equation}
where in the last step we made use of the relation (\ref{CvarM}) between the
variations $\delta I$ and $\delta M$ and inserted the explicit
results (\ref{Cintmu1}) for $I_c = I(1)$ and $M_c = M(1)$.
Using the result (\ref{CvarI}) for $\delta I/I_c$ to
lowest order we obtain
\begin{equation}
\delta \kappa  \; \; = \; \; - \; \varepsilon
\ln (\varepsilon) \; \sigma \; ,
\label{Cvar2}
\end{equation}
where we introduced the abbreviation
\begin{equation}
\sigma \; \equiv \; \frac{1}{8} \left[- \frac{\gamma \pi^2}{256} +
\frac{4 \lambda^c}{3} \right] \; .
\label{Csigma}
\end{equation}
Inverting equation (\ref{Cvar1}) (again taking into account only the
leading order) gives
\begin{equation}
\varepsilon \; = \; - \; \sigma^{-1} \delta \kappa \;
\left[ \ln \left( \sigma^{-1} \delta \kappa
\right) \right]^{-1}\; .
\label{Cinv}
\end{equation}
This equation is the relation between the variation $\delta
\kappa$ and $\varepsilon$ which is implied by the necessary
condition (\ref{Cnecc}). In the final step we expand the derivative of the
free energy (\ref{Cdfmu}) at
$\mu = 1$ and use the result (\ref{Cinv}) to obtain the variation of the
derivative in terms of $\delta \kappa $. Expanding
(\ref{Cdfmu}) gives
\begin{equation}
\frac{d  f}{d \kappa } \; \; \; = \; \; \; -2 \frac{M_c}{I_c}
\left[ 1 + \frac{\delta M}{Mc} - \frac{\delta I}{I_c} \right]
\; \; \; = \; \; \;
-\frac{2}{3} \; - \; \frac{1}{3} \varepsilon \ln(\varepsilon)
\; - \; \varepsilon \; .
\label{Cfin1}
\end{equation}
Using (\ref{Cinv}) we obtain
\begin{equation}
\frac{d  f}{d \kappa} \; \; \; = \; \; \;
-\frac{2}{3} \; \; + \; \; \frac{1}{3} \sigma^{-1} \delta
\kappa \; \; + \; \;
\sigma^{-1} \delta \kappa  \; \left[ \ln \left( 
\sigma^{-1} \delta \kappa \right) \right]^{-1} \; .
\label{Cfin2}
\end{equation}
We remark, that the case $\mu < 1$ with expansion $\mu = 1 - \varepsilon,
\; \varepsilon > 0$ changes only the sign of the argument of the
logarithm. Differentiating the last result with respect to $\delta \kappa$
establishes the singular behaviour of the third derivative of the free
energy with respect to $\kappa$. Thus we find a third order phase transition
for $\mu = 1$. The critical line is a straight line given by
(\ref{Ccritlin}).

It is important to notice, that at (see Equation (\ref{Cvar1}))
\begin{equation}
\frac{\lambda^c}{\gamma} \; =  \;\frac{3 \pi^2}{1024} \; ,
\label{Clamterm}
\end{equation}
the leading term in the
expression for $\delta \kappa$ vanishes. Equation (\ref{Cvar1})
is reduced to the simpler relation
\begin{equation}
\frac{\delta \kappa }{\gamma} \; = \; - \varepsilon \;
\frac{\lambda^c}{2 \gamma} \; .
\label{Cvar3}
\end{equation}
At this point the expansion of $df/d\kappa$ gives
\begin{equation}
\frac{d f}{d \kappa } \; = \; -\frac{2}{3} \; - \;
\frac{1}{3} \varepsilon \ln(\varepsilon) \; - \; \varepsilon \; = \;
-\frac{2}{3} \; + \; \frac{2}{3 \lambda^c}  \delta \kappa
\ln\left(- \frac{2}{\lambda^c} \delta \kappa \right) \; + \;
\frac{2}{\lambda^2} \delta \kappa  \; .
\label{Cfinscnd}
\end{equation}
Again the case $\mu < 1$ differs only by the sign of the argument of the logarithm. Differentiation with respect to $\delta \kappa$
shows, that the phase transition has turned to second order at that point.
Using (\ref{Ccritlin}) one can compute also the $\kappa/\gamma$
coordinate of the second order point giving $\kappa/\gamma = \pi^2/1024,
\lambda/\gamma = 3 \pi^2/1024$. In fact the more global analysis in the
next section will show, that the third order line terminates at
the second order point $\kappa/\gamma = \pi^2/1024,
\lambda/\gamma = 3 \pi^2/1024$, and continues as a first order line.

\subsection{Regime of the first order phase transition}

\begin{figure}
\epsfysize=3in
\epsfbox[0 455 520 717] {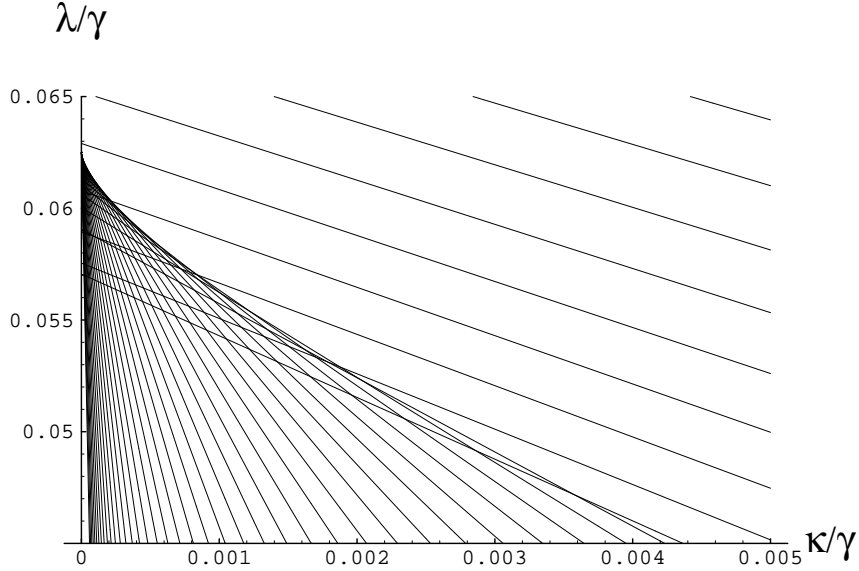}
%\vskip5in
%\special{eps:lines.ps x=5in y=4in}
\caption{{\it Plot of the lines (3.18) for $\mu$ ranging from 
$0.4$ (upper right corner) to $75$ (line at the extreme left).  
The region of overlapping lines corresponds to
a region of first order phase transition. \label{linesfig} }}
\smallskip
\end{figure}

As pointed out at the end of the last section at the point $\kappa/\gamma
= \pi^2 / 1024$, $\lambda/\gamma = 3 \pi^2 / 1024$, the third order
transition along the $\mu=1$ line (\ref{Ccritlin}) changes to second order.
This unusual behaviour requires further investigation which we will
carry out in this section. To begin, a graphical analysis
of the phase diagram is most useful and in Figure
\ref{linesfig} we plot a number
of representatives of the family of lines (\ref{Cnecc}) for a range of
values of $\mu$.
It is clear that in most of the $\kappa, \lambda$-plane
points are in a one-to-one correspondence with values of the parameter $\mu$.
This correspondence breaks down though in a small
region near the $\lambda / \gamma$ axis between $\lambda / \gamma =0.05$
and $\lambda / \gamma =0.0625$.
Due to the behaviour of the slope and intercept in the linear equation
(\ref{Ccritlin}) lines begin to overlap for increasing $\mu$ starting at
$\mu \sim 1$ and continuing as $\mu \rightarrow \infty$.
In this overlap region the phase diagram is folded at the vertex
$\kappa/\gamma= \pi^2 / 1024$, $\lambda/\gamma = 3 \pi^2 / 1024$, and
each point falls on three  different
lines of constant $\mu$. Consequently the system simultaneously
admits three configurations with different free energies in
this region of the phase space. This circumstance
allows for a first order phase transition to develop along a line where the
free energies of the different phases are equal.

The edges of the triangular  first order region in Figure \ref{linesfig}
is given by a caustic of lines from the
one parameter family (\ref{Cnecc}).  The boundary is defined by the curve where
the family of curves is stationary with respect to $\mu$.  This condition can
be used with (\ref{Cnecc}) to give a definition of the boundary caustic.
The stationary condition can be solved with the parametric result
\beq
\frac{\kappa}{\gamma} \; \; = \; \;
\frac{1}{4 I^2(\mu) } \; \frac{ I^{\prime}(\mu) M(\mu) +
M^{\prime}(\mu) I(\mu) }{M^{\prime}(\mu) I(\mu) -I^{\prime}(\mu) M(\mu) }
\label{caustic} \; .
\eeq
As can be seen, the curve given by (\ref{caustic}) intersects the
$\lambda / \gamma$
axis at two points: $0.057024~ (\mu = 0.95324)$ and $1/16~(\mu= \infty)$
and reaches a singular maximum in the $\kappa / \gamma$ direction
for $\mu = 1$ at the point $\kappa / \gamma= \pi^2 /1024$.  The end of
this region of first order transitions agrees with the position of the second
order transition point which was determined by the analysis of critical behaviour in the previous section.

\begin{figure}
\epsfysize=3in
\epsfbox[0 438 545 720] {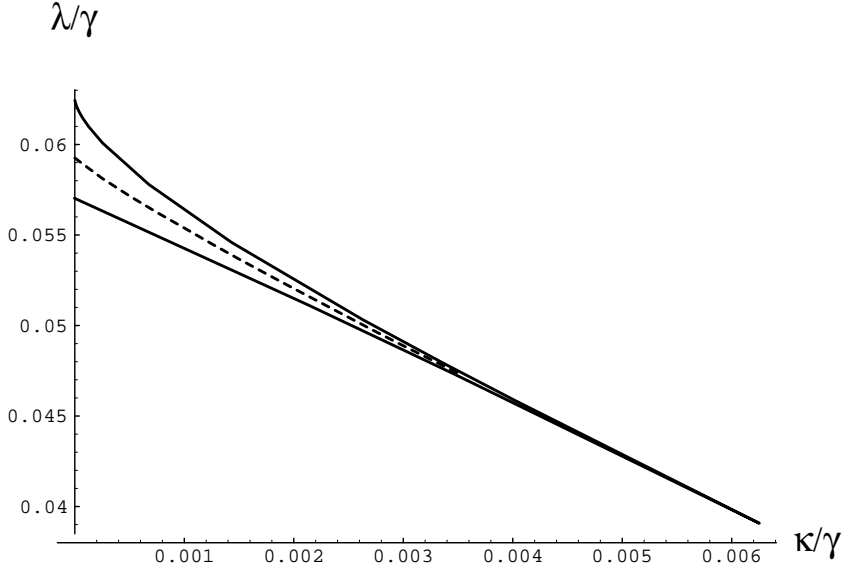}
\caption{ {\it Plot of the boundary of the multiple phase region.
The boundary (solid curve) is given by a caustic of lines in the one-parameter
family (3.18). The dotted curve shows the numerically determined 
first order line.
\label{causticfig} }}
\end{figure}

Once one has determined the region where different phases can co-exist the
next issue to address is that of the position of the line of first order
phase transitions where different phases have the same free energy.
This line can be determined for given $\kappa / \gamma$ by the simultaneous
solution for the parameters $\mu_1$ and $\mu_2$ of the pair of equations
\beq
\frac{I(\mu_2) M(\mu_2) - I(\mu_1) M(\mu_1)}{ 4 \; I(\mu_1) I(\mu_2)} \; =
\; \frac{ \kappa}{\gamma} \Big[
M(\mu_2) I(\mu_1) - I(\mu_2) M(\mu_1) \Big] \; ,
\label{firsto1}
\eeq
and
\beq
\frac{1}{6} \left( \frac{\mu_1}{I^2(\mu_1)} -
\frac{ \mu_2}{I^2(\mu_2)} \right) -\frac{1}{12}
\left( \frac{M(\mu_1)}{I^3(\mu_1)} - \frac{ M(\mu_2)}{I^3(\mu_2)} \right)
\; = \;
\frac{\kappa}{\gamma} \frac{I(\mu_2) M(\mu_1) - M(\mu_2) I(\mu_1)}
{I(\mu_1) I(\mu_2)} \; .
\label{firsto2}
\eeq

\begin{figure}
\epsfysize=3in
\epsfbox[50 465 540 700] {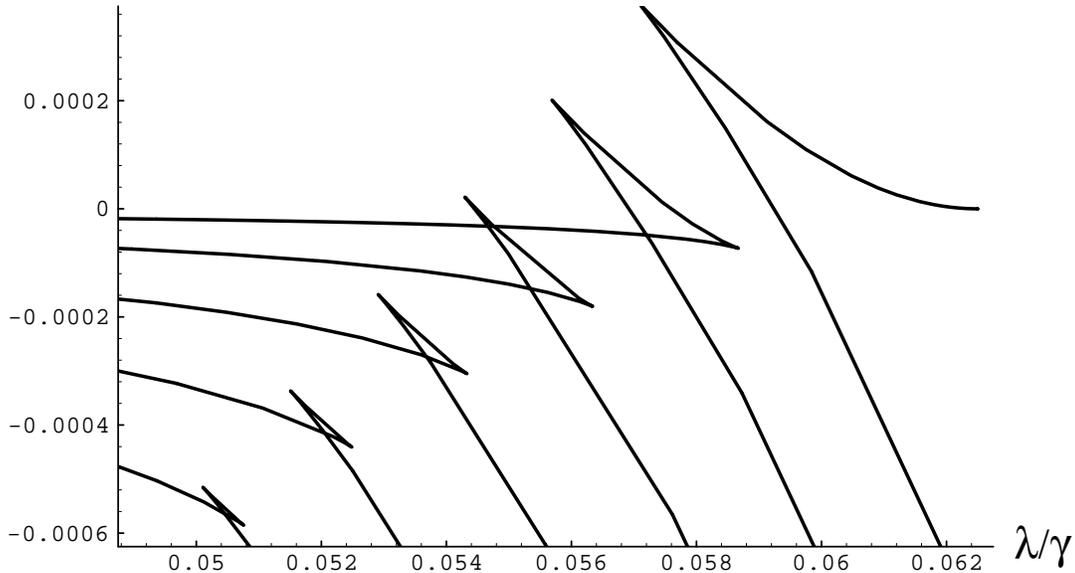}
\caption{{\it Free energy $f/\gamma$ as a function 
of $\lambda/\gamma$ in the region of first order phase transitions.
Each curve is plotted for fixed $\kappa / \gamma$ which from right to left
is $\kappa / \gamma  =
0,~0.0005,~0.001,~0.0015,~0.002,~0.0025$ \label{freeE}} }
\end{figure}

Unfortunately, these equations are analytically intractable.  Again we
turn to a graphical analysis to gain further insight.  In Figure \ref{freeE}
we plot the free energy of the system as a function of $\lambda / \gamma$
for different values of fixed $\kappa / \gamma$.  From here it is easy to
see a number of features of the region of first order transitions.
Increasing $\mu$
traverses these curves in a clock-wise rotation so that free energy increases
for small values of $\mu$, intersecting the nearly horizontal
large $\mu$ free energy.  This intersection point is a graphical
demonstration of the first order transition which occurs here
as the model jumps from weak $(\mu<1)$ to strong coupling $(\mu$ large$)$.
Each phase continues to exist after the transition point and may be
reached by an adiabatic process until ending in cusps which mark the
boundaries of the first order region in the $\lambda / \gamma$ axis.
It is interesting to note that there is an energetically unfeasible
intermediate ``medium coupling'' phase which connects the weak and strong
phases. Hence, for fixed $\kappa / \gamma$ there exist three distinct
configurations of the system for given $\lambda / \gamma$ in the
region of first order transitions.

\section{Discussion}
We show that the whole phase diagram can be understood in terms
of the energy densities of the fundamental charges the adjoint charges and
the flux lines. Higher winding Polyakov loop operators will be used to 
characterize the phases of the model.

\setcounter{equation}{0}
\subsection{Energy densities}
In this section we identify the expressions which should be interpreted
as the energy densities of the fundamental charges
the adjoint charges and the flux lines.

Using the definition (\ref{CfreeE}) for the leading ($O(N^2)$) coefficient
$f$ of the density of the free energy we can
formally denote the partition function $Z[T/e^2,\lambda_n, \kappa_n]$,
defined in (\ref{partition}), (\ref{Seff0}) as (note that $\gamma$ is
related to $T/e^2$ via (\ref{gammadefn}))
\begin{equation}
Z[ T/e^2, \lambda_n, \kappa_n ] \; \sim \; \exp \left( -N^2 L \;
f(\gamma) \right) \; .
\label{Zf}
\end{equation}
$L$ denotes the space-like extension which is infinite. Thus (\ref{Zf})
as it stands can only be understood formally, but could be made an
exact identity by putting the system in a finite box. However, we show
that derivatives of $f$ have an interpretation as energy densities also in
the infinite system.

Using (\ref{partition}), (\ref{Seff0}) we obtain
\[
\frac{d \ln Z}{d \; T^{-1}} \; \; = \; \;- \langle H_{Flux} \rangle 
\; \; =  \; \;
- N^2 L \; \frac{df}{d \; T^{-1}} \; ,
\]
where in the last step we made use of (\ref{Zf}). Using the definition
for $\gamma$ (\ref{gammadefn}) we obtain
\begin{equation}
\frac{df}{d\gamma} \; \;  =  \; \; \frac{1}{N^2}
\frac{\langle H_{Flux} \rangle}{L} \Big/ e^2 \frac{N}{2} \; \;
\equiv \; \; \rho_{Flux} \; .
\label{rfluxdef}
\end{equation}
$\rho_{Flux}$ is the leading coefficient ($O(N^2)$) of the contribution of
the electric flux to the energy density, measured in units of $e^2 N/2$.
Note that $e^2 N/2$ is a proper unit for energy densities, since it is
invariant in the limit $N \rightarrow \infty$ (compare (\ref{gammadefn})).
The inclusion of the factor 1/2 is natural, since the quadratic Casimir
operator behaves (for $U(N)$) as $N/2$ at large $N$.

In the same way we can proceed to identify the contribution of the
adjoint particles to the energy density\footnote{Again we restrict ourselves
to only one (real) pair $\lambda_J, \kappa_J \neq 0$, although some of the
results of this section can be immediately taken over to the case
of several nonvanishing couplings.}
\begin{equation}
- \frac{\lambda_J}{\gamma}
\frac{df}{d\lambda_J} \; \; = \; \; \frac{1}{N^2}
\frac{\langle H_{Ad} \rangle}{L} \Big/ e^2\frac{N}{2} \;
\equiv \; \; \rho_{Ad} \; \;.
\label{raddef}
\end{equation}
Finally for the contribution of the fundamental charges and their conjugates
we obtain
\begin{equation}
- \frac{\kappa_J}{\gamma}
\frac{df}{d\kappa_J} \; \; = \; \; \frac{1}{N^2}
\frac{\langle H_{F,\bar{F}} \rangle}{L} \Big/ e^2\frac{N}{2} \; \; \equiv
\; \; \rho_{F,\bar{F}} \; .
\label{rfundef}
\end{equation}
The differential equation (\ref{Cdiffe}) gives
\begin{equation}
\rho_{Flux} - \rho_{Ad} - \rho_{F,\bar{F}} \; \; = \; \; \frac{1}{\gamma} 
\; f \; \; =  \;  \; \frac{1}{N^2} \frac{F}{L} \Big/ e^2 \frac{N}{2} \; ,
\label{rhodiffe}
\end{equation}
where $F$ denotes the free energy of the system.
Thus the density of the free energy is the difference of the
contribution of the flux and the contribution of the charges.

When interpreting the phase diagram it is convenient to express the
densities in terms of $\kappa/\gamma, \lambda/\gamma, I(\mu)$ and
$M(\mu)$. Using (\ref{Cdfdl}), (\ref{Cfmu}), (\ref{Cdfmu}) and
(\ref{rhodiffe}) we obtain
\begin{eqnarray}
\rho_{Flux} \; &=& \; \frac{1}{6I(\mu)^2} \left[ \mu +
\frac{M(\mu)}{I(\mu)} \right] \; - \frac{1}{96} \; , \label{rmflux}\\
\rho_{Ad} \; &=& \; \frac{\lambda}{\gamma} \;
\left( \frac{M(\mu)}{I(\mu)} \right)^2 \; , \label{rmad} \\
\rho_{F,\bar{F}} \; &=& \; 2 \frac{\kappa}{\gamma}
\frac{M(\mu)}{I(\mu)} \; . \label{rmfun}
\end{eqnarray}
The fact that $\rho_{Flux}$ does not depend on $\kappa/\gamma$ and
$\lambda/\gamma$ is most remarkable. It proves that the lines of constant
$\mu$ are the lines of constant flux. In the last section we showed
that the lines of constant $\mu$ are responsible for the 
structure of the phase
diagram. This implies that the whole phase diagram can be
understood by the abundance of flux.

We remark that the necessary condition (\ref{Cnecc}) 
gives rise to another quantity which is constant on lines with fixed $\mu$.
It is the linear combination
\begin{equation}
\rho_{F,\bar{F}} \; + \; 2 \rho_{Ad} \; = \;
\frac{1}{2} \frac{M(\mu)}{I(\mu)^3} \; .
\label{lincomb}
\end{equation}

\subsection{Higher windings of Polyakov loop operators}
In addition to analyzing the phase diagram in terms 
of contributions of the electric flux and adjoint and fundamental
charges to the free energy we also investigate
higher windings of the Polyakov loop operators. 
As we will now demonstrate, their behaviour changes fundamentally 
at the line $\mu=1$. It can be used to characterize
the difference between the regions $\mu>1$ and $\mu<1$ and
the physics in these different phases.

In the model with only the pair $\kappa_J, \lambda_J \neq 0$ we will be 
concerned with the quantities
\begin{equation}
P_k(\mu) \; \; \equiv \; \; \lim_{N \rightarrow \infty} \frac{1}{N} 
< \Tr g^{kJ}> \; \; = \; \;  c_{kJ} \; . 
\label{poldef}
\end{equation}
Using (\ref{cnphi}), (\ref{Cname}) and (\ref{CEscal}) we obtain
\begin{equation}
c_{kJ}e^{i\beta kJ} \; = \; \frac{M_k(\mu)}{I(\mu)}  \; ,
\label{ckj}
\end{equation}
where we have defined
\begin{equation}
M_k(\mu) \; \equiv \; \frac{2}{\pi} \int_{-\pi}^{\pi} \cos{k \theta}
\sqrt{\mu + \cos(\theta)} \;
\mbox{H}(\mu + \cos(\theta)) \; d\theta \; .
\label{kmoment}
\end{equation}
Note that as in the cases of $I(\mu)$ and $M(\mu)$ (see Appendix A.2), 
we have performed a transformation of the
integration variable which removes the dependence on $J$.
The most important thing to note about this last equation is that
these higher windings depend only on the parameter $\mu$ and
hence give a unified picture of different regions of the
phase space as does the electric flux density $\rho_{Flux}$.

In Appendix A.3 we show that the $M_k(\mu)$ and thus the higher windings 
of the Polyakov loop operators $P_k(\mu)$ undergo a drastic change of
character at the point $\mu=1$. We find,
that for $\mu > 1$ they are exponentially suppressed with 
increasing $k$, while for $0 \leq \mu \leq 1$ there is only 
power law decay. Our results
are (use (\ref{Aexpsup}), (\ref{mom1}))
\begin{equation}
|P_k(\mu)| \; \leq \; \Big(r(\mu)\Big)^k \; \frac{1}{2}
\sqrt{2\mu + r(\mu) + r(\mu)^{-1}} \; \; \; \; \; \; \; 
\mbox{for} \; \mu > 1 \; ,
\label{Pexpsup}
\end{equation}
with $r(\mu) \equiv (1 + \mu - \sqrt{\mu^2 -1})/2 < 1$ for $\mu > 1$.
For $\mu = 1$ we obtain
\begin{equation}
|P_k(\mu)| \; = \; \frac{1}{4k^2 - 1} \; ,
\label{P1}
\end{equation}
and for $-1 \leq \mu < 1$ it is shown in Appendix 3.1 
that exponential suppression is excluded.

In terms of a Fourier decomposition of the 
eigenvalue distribution $\rho_0(\theta)$, this result is a comment
on the smoothness of the distribution.  For $\mu>1$, $\rho_0(\theta)$
has support on the entire circle and is infinitely differentiable,
whereas for $\mu<1$ the zeros of $\rho_0(\theta)$ with their infinite
slope contribute higher frequencies to the Fourier series.

Physically one would like to be able to associate the change
in behaviour of the higher winding Polyakov loops to a change in 
character of the gas of fundamental and 
adjoint charges.  As seen above, and as will be expanded on 
in the next section, the phase diagram may be broadly divided 
into two regions.  We have proven that  $\mu=1$
denotes a line of third order phase transitions up to a point 
where a first order line develops and there is a competition 
between phases with $ \mu <1 $ and $ \mu >1 $. Thus the behaviour 
of the $P_k(\mu)$ is different on both sides of the third order
section of the critical line, as well as on the two sides of
the first order section. 

It is interesting to notice, that the large-$k$ behaviour of the 
higher winding Polyakov loop operators $P_k(\mu)$ can also be 
interpreted as the long distance behaviour of a gauge invariant 
two-point function. In this picture $P_k(\mu)$ then 
corresponds to a Polyakov loop winding $k$-times around compactified time 
of the Euclidean model (compare (\ref{ftpi})). Thus the $P_k(\mu)$ can
be considered as the expectation value of a gauge invariant two-point
function with time-like separation $k/T, \; k = 1,2,3, ...$. For other 
models it was already proposed earlier \cite{bf}, 
to use the changing decay properties of such operators as a confinement 
criterion. 

It remains an interesting open question if the change in behaviour 
of the higher winding Polyakov loops persists in higher dimensions and 
for finite $N$. If it does, it would provide a powerful tool to study 
deconfinement transitions, in particular in lattice simulations,
where the $P_k(\mu)$ are much simpler to analyze than other criteria,
as e.g. the area law of the Wilson loop.
\subsection{Physical interpretation of the phase diagram}

\begin{figure}
\epsfysize=3in
\epsfbox[-215 315 385 715] {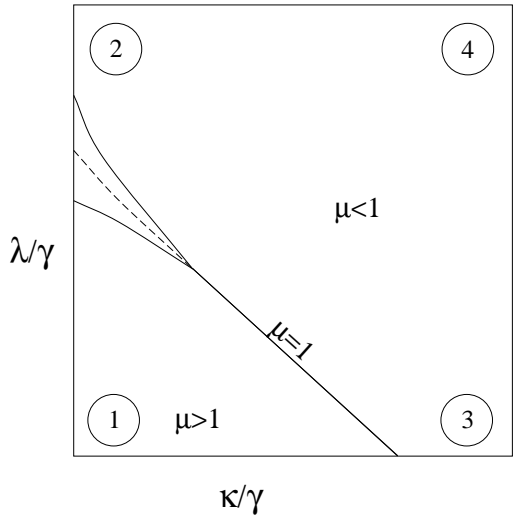}
\caption{{\it Schematic picture of the phase diagram. 
The doted curve marks the first order 
part of the critical line. The solid curves above and below it are the 
boundaries of the area with two possible phases. They join at a point
which shows second order behaviour. For larger $\kappa/\gamma$
we find a third order line ($\mu = 1$) marked by a solid line. The numbers
label four extremal corners of the phase diagram, which we discuss below.
We also indicate the range of the auxiliary parameter $\mu$.
\label{phasediag}} }
\end{figure}

Figure \ref{phasediag} gives a schematic plot of the phase diagram.
In the following we  explore the physical behaviour of the
system in various regions of the phase diagram.
\vskip3mm
\noindent
{\it The line $\kappa = 0$ (purely adjoint model)}
\vskip1mm
\noindent
In the case of only adjoint charges
(all $k_n = 0$) the action is invariant under transformations in the center
of the gauge group, $g \rightarrow z g$ (compare (\ref{zsym})). When this symmetry is faithfully represented we expect
\begin{equation}
c_n(x) \; = \; \lim_{N \rightarrow \infty} \frac{1}{N} \langle {\rm
Tr}~g^n(x) \rangle \; = \; 0 \; \; \; \; \; \; \forall n \neq 0 \; ,
\label {consol}
\end{equation}
since
\[
{\rm
Tr}~g^n(x)~\rightarrow z^n~{\rm Tr}~g^n(x) \; .
\] From (\ref{coll1}), (\ref{coll2}) it is obvious that (\ref{consol})
which corresponds to constant eigenvalue density $\rho \equiv 1/2\pi$ 
is a solution. The stability
analysis performed in \cite{stz} shows that this solution ("confined
phase") is stable for $\lambda/\gamma \in [0, 1/16]$. Since the upper
boundary of the multiple phase region
is at $\lambda/\gamma = 1/16$ we
confirm this result. The confined phase at $\kappa/\gamma = 0$ corresponds
to $\mu = \infty$ (compare Fig. \ref{linesfig}). Inserting the results
(\ref{largemu}) for the large-$\mu$ expansion in (\ref{rmflux}), 
(\ref{rmad}) and (\ref{Cfmu}) respectively, we find energy densities
$\rho_{Flux} = 0$, $\rho_{Ad} = 0$ and $f= 0$ 
in the confined phase of the purely adjoint model. 
Also the higher winding Polyakov loops $P_k(\mu)$ vanish as $\mu \rightarrow 
\infty$. All this supports the picture of adjoint charges being 
bound pair-wise in singlets under center transformations. The phenomenon of
adjoint charges bound in such ``hadrons'' is called confinement. 

For higher $\lambda/\gamma$ the center symmetry is spontaneously broken.
This corresponds to a solution, where the coefficients $c_n$ are
non-vanishing as well as the densities $\rho_{Flux},
\rho_{Ad}$ and the energy density $f$. Typical values (taken at
$\lambda/\gamma = 0.07$) are, $\rho_{Flux} = 0.017714$, 
$\rho_{Ad} = 0.020385$, $f/\gamma = -0.002671$.
This ``deconfining phase'' has a
nonvanishing overlap with the confining phase and gives rise to a
phase transition which is of first order. The corresponding
critical coupling can be computed numerically giving 
$\lambda/\gamma = 0.059250$. This result is in agreement with the 
value given in \cite{stz}.
\vskip5mm
\noindent
{\it The line $\lambda = 0$ (purely fundamental model)}
\vskip1mm
\noindent
The center symmetry of the action is explicitly broken when there are
fundamental charges (some $\kappa_n \neq 0$). In particular this is the 
case when we consider the model with only fundamental charges (and their
conjugate). Thus we cannot expect to find a confining solution with vanishing 
$c_n$ which would correspond to vanishing densities $\rho_{Flux}, \;
\rho_{F,\bar{F}}$.

For the purely fundamental model we have
established the existence of a third order 
phase transition at $\kappa/\gamma =  \pi^2/512$ 
(set $\mu=1, \lambda = 0$ in (\ref{Cnecc})). 
The phase with $\kappa/\gamma$ smaller than the critical 
value is characterized by low densities $\rho_{Flux}, \; \rho_{F,\bar{F}}$
and exponential suppression of the higher winding Polyakov loop operators
$P_k(\mu)$, while for $\kappa/\gamma$ greater than the critical value we 
find higher densities and power-like suppression of the $P_k(\mu)$. 

Using the expansion (\ref{largemu}) for large $\mu$, one can compute the 
behaviour of the densities at small $\kappa/\gamma$. We obtain (expand
(\ref{Cnecc}) for large $\mu$, to find the relation between $\kappa/\gamma$
and $\mu$)
\[
\rho_{Flux} \; = \; 16 \left( \frac{\kappa}{\gamma} \right)^2 
\; + \; O \Big((\kappa/\gamma)^4 \Big)\; \; \; \; , 
\; \; \; \; 
\rho_{F,\bar{F}} \; = \; 32 \left(\frac{\kappa}{\gamma} \right)^2 
\; + \; O \Big((\kappa/\gamma)^4 \Big) \; .
\]
It is interesting to notice, that the two densities vanish with fixed 
ratio 2. This could be interpreted as a dilute gas of "hadrons" 
formed of pairs of fundamental charges
and their conjugates bound together by flux. 
\vskip5mm
\noindent
{\it The third order line}
\vskip1mm
\noindent
In Section 3.3 we established that at $\mu = 1$ there is a 
third order phase transition. For $\lambda/\gamma < 3\pi^2/1024$ the critical
line is given by Equation (\ref{Ccritlin}). The third order behaviour
terminates at $\kappa/\gamma = \pi^2/1024,
\lambda/\gamma = 3\pi^2/1024$ becoming second order at that point,
continuing as a line of first order transition.
Using the values (\ref{Cintmu1}) we obtain
for the energy densities of adjoint and fundamental charges
\[
\rho_{F,\bar{F}} \; = \; \frac{2}{3} \frac{\kappa}{\gamma}  \; \; \; \; ,
\; \; \; \; \rho_{Ad} \; = \; \frac{1}{9} \frac{\lambda}{\gamma}  \; .
\]
The energy density of the flux is constant along the third order line
and given by (evaluating (\ref{rmflux}) at $\mu = 1$)
\[
\rho_{Flux} \; = \;  0.006718 \; .
\]
The invariant linear combination (\ref{lincomb}) assumes the value
\[
\rho_{F,\bar{F}} \; + \; 2 \rho_{Ad} \; = \; 0.012851 \; ,
\]
on the critical line. 

The two phases above and below the third order line differ by the 
values of the energy densities, with higher values of density 
for large $\kappa/\gamma$ and $\lambda/\gamma$.
In particular the flux-density is higher for $\mu < 1$, and lower for
$\mu > 1$. Thus we call the phase with $\mu > 1$ the ``low density phase'',
while the other phase will be referred to as ``high density phase''. 
Since the energy density of the flux $\rho_{Flux}$ vanishes entirely
only for the purely adjoint model discussed in the last paragraph, we reserve
the term ``confined/deconfining phase'' for this case. 

The most dramatic change is seen in the behaviour of the higher 
winding Polyakov loop operators $P_k(\mu)$. We find exponential suppression 
with increasing $k$ for $\mu >1$ which corresponds to the low density phase 
(small $\kappa/\gamma, \lambda/\gamma$). For $\mu < 1$ (high density phase) 
we showed that the $P_k(\mu)$ decay only power-like with $k$. On the critical
line ($\mu = 1$) we find $|P_k(\mu)| = 1/(4k^2 -1)$. Together with the 
behaviour of the flux density $\rho_{Flux}$, this allows to understand 
the whole phase diagram by the abundance of flux. 
\vskip5mm
\noindent
{\it The first order line}
\vskip1mm
\noindent
The first order line comes about through the coexistence of two phases with
equal free energy (compare Figure \ref{freeE}). The region where the
phases coexist is shown in Figure \ref{causticfig}. 
The two phases can be distinguished by the different
contributions of flux and sources to the
energy density (high density and low density phase). 
Typical values are (taken at $\kappa/\gamma =0.002,
\lambda/\gamma = 0.05203$)
\[
\rho_{Flux} \; = \; 0.003328  \; \; \; ,  \; \; \;
\rho_{Ad} \; = \; 0.002811 \; \; \; ,  \; \; \;
\rho_{F,\bar{F}} \; = 0.000929 \; ,
\]
for the low density phase which connects to the confined phase of the 
purely adjoint model ($\kappa/\gamma = 0, \lambda/\gamma \in [0,1/16]$) and
\[
\rho_{Flux} \; = \; 0.008999 \; \; \; , \; \; \;
\rho_{Ad} \; = \; 0.007856 \; \; \; , \; \; \;
\rho_{F,\bar{F}} \; = 0.001554 \; ,
\]
for the high density phase which connects to the deconfining phase of 
the purely
adjoint model ($\kappa/\gamma = 0, \lambda/\gamma > 0.057024)$. Although
unlike for the purely adjoint case, for $\kappa/\gamma > 0$
we have no immediate interpretation using symmetries, since for 
$\kappa/\gamma > 0$ the center symmetry is always broken explicitly.
However, the physical behaviour in the first order region is still 
reminiscent of the purely adjoint case. The values of all three densities 
are considerably enhanced in the high density phase. Since the numbers we 
quoted are taken for a point on the critical line, they give rise to the
same free energy (Note that the free energy is 
not the sum of the three contributions, but rather related to them by 
(\ref{rhodiffe}).). 

It is important to notice, that also for the first order 
section of the critical line
the higher winding Polyakov loops provide a criterion to distinguish the
two phases. At the first order line a phase with $\mu >1$ and a phase 
with $\mu < 1$ coexist. Thus also for the first order transition
we find the characteristic exponential decay in the low density phase, but 
power-law behaviour in the high density phase.

\vskip5mm
\noindent
{\it The second order point}
\vskip1mm
\noindent
In the end of Section 3.3 it was established that the point
$\kappa/\gamma = \pi^2/1024, \lambda/\gamma = 3\pi^2/1024$,
which separates the third order part of the critical line from
the first order part, is of second order. At this point the energy
density for the adjoint and fundamental charges
respectively are given by (insert $\mu = 1$ and the coordinates of
the second order point in (\ref{rmad}), (\ref{rmfun}))
\[
\rho_{Ad} \; = \; \frac{\pi^2}{3072} \; \; \; \; , \; \; \; \;
\rho_{F,\bar{F}} \; = \frac{\pi^2}{1536} \; \; \; \; , \; \; \; \;
\frac{\rho_{F,\bar{F}}}{\rho_{Ad}} \; = \; 2  \; .
\]
If one decreases $\kappa/\gamma$ from its value at the second order
point, the adjoint charges start to dominate, and the ratio
$\rho_{F,\bar{F}}/\rho_{Ad}$ decreases - the first order region
emerges. Conversely, with increasing $\kappa/\gamma$ the fundamental
charges start to dominate the system and
$\rho_{F,\bar{F}}/\rho_{Ad}$ increases.

The existence of the phase transitions which we have discussed in this
section is a bit surprising, considering the low dimensionality of the
system.  In fact, if we first consider the model with purely adjoint
quarks as discussed in \cite{stz}, there is a discrete symmetry
under transformations in the center of the gauge group and the system is
one-dimensional.  Such a system cannot have a phase transition if $N$ is
finite.  When $N$ is infinite, the center symmetry becomes $U(1)$ and the
group index behaves like another dimension, so the effective
dimensionality of the model is two.  In this case, the symmetry can be
spontaneously broken (and there can be deconfinement) because the
interaction potential has infinite range in eigenvalue space.  This is
also seen by considering the translation mode in eigenvalue space, which
is the derivative by $\alpha$ of the eigenvalue density $\rho$.  
When the support
of the density is a subset of the interval $[-\pi,\pi]$ the derivative of
$\rho$ is not a normalizable wavefunction (divergence in 
the norm coming from the edge of 
the distribution) for the zero mode and thus it is not
effective in restoring the symmetry.  Similarly, we expect that, when we
turn on the density of fundamental representation quarks, the first or
third order phase transition happens only at infinite $N$.

One can think of the infinite $N$ limit here as something analogous to the
infinite volume limit in the case of a ferro-magnetic system.  In the
latter, a phase transition in the mathematical sense only occurs when the
volume is infinite.  However, in the physically relevant case where the
volume is large but finite, the phase transition is clearly seen.
Similarly, in the present case of large $N$, the phase transition only
occurs in the strict mathematical sense in the large $N$ limit.  It should
nevertheless be visible when $N$ is finite as long as finite size effects
are not too large.  One would expect that the tunneling which restores
symmetry at finite $N$ to be of order $\exp(-N^2\cdot{\rm const.})$ which
could be very small even for rather small values of $N$.  It would be very
interesting to check this idea using lattice Monte Carlo simulations of
these systems.
\vskip5mm
\noindent
{\it Extremal corners of the phase diagram}
\vskip1mm
\noindent
Finally we discuss extremal corners of the phase diagram, which
are labeled 1,..,4 in Figure \ref{phasediag}. Table
1 gives the values of $\kappa/\gamma, \lambda/\gamma$
and the values of the contributions to the energy density.
Obviously the four extremal cases can easily be understood by the
magnitude of the energy densities $\rho_{Flux}$, $\rho_{Ad}$ and
$\rho_{F, \bar{F}}$. Point 1 is in the extremal corner of the 
low density phase. 
All three densities are rather small. Points 2 and 3 are both in the
high density phase, in areas which are dominated by adjoint charges
(Point 2) and fundamental charges (Point 3). It is nice to see,
how the energy density of the sources is dominated by the contributions 
of the adjoint charges and fundamental charges respectively. Finally
Point 4 is in a region where both the density of the adjoint charges
as well as the fundamental charges is high and of the same magnitude.
\begin{center}
\begin{tabular}{|c||c|c|c|c|}
\hline
 & {\bf 1} & {\bf 2} & {\bf 3} & {\bf 4} \\
\hline \hline
$\kappa/\gamma$ & $0.00025$ & $0.00025$ & $0.25$ & $0.25$ \\
$\lambda/\gamma$ & $0.000252$ & $0.25$ & $0.000251$ & $0.25$\\
\hline
$\rho_{Flux}$ & $1.033\times 10^{-6}$ & $0.063260$&$0.072505$&
	 $0.105058$ \\
$\rho_{Ad}$ &$4.064 \times 10^{-9}$& $0.160997$&$0.000170$ &
	$0.189518$\\
$\rho_{F,\bar{F}}$&$2.008 \times 10^{-6}$&$0.000401$&
		$0.411543$&$0.435337$ \\
\hline
\end{tabular}
\vskip 3mm
Table 1: {\it Values of $\kappa/\gamma, \lambda/\gamma$ and the
contributions to the energy \\ densities
at four characteristic points in the phase diagram.}
\end{center}

It is interesting to notice, that when pushing the Points 2,3 and 4 to even 
higher values of $\kappa/\gamma$ and $\lambda/\gamma$, one finds that
the contribution of the 
charges to the energy density grows faster than the contribution of the flux. 
These extremal areas correspond to $\mu \sim -1 + \varepsilon$. Expanding
the necessary condition (\ref{Cnecc}) and the densities (\ref{rmflux}) - 
(\ref{rmfun}) for $\mu \sim -1 + \varepsilon$ (use (\ref{musim1}))
we obtain
\begin{eqnarray}
\rho_{Flux} \; = \;  \left( \sqrt{\frac{\kappa}{\gamma}
+ \frac{\lambda}{\gamma} } \; \frac{\sqrt{2}}{6} - \frac{1}{96} \right) \; 
\Big[ 1 + o\Big( \sqrt{ \kappa/\gamma + \lambda/\gamma }\Big) \Big] \; , \nonumber \\
\rho_{Ad} \; = \; \frac{\lambda}{\gamma} \; 
\Big[ 1 + o\Big( \sqrt{ \kappa/\gamma + \lambda/\gamma }\Big) \Big]
\; \; \; \; , \; \; \; \;
\rho_{F,\bar{F}} \; = \; \frac{\kappa}{\gamma} \; 2 
\Big[ 1 + o\Big( \sqrt{ \kappa/\gamma + \lambda/\gamma }\Big) \Big]
\; ,
\nonumber
\end{eqnarray}
in the limit $\kappa/\gamma, \lambda/\gamma \gg 1$. This shows that at 
very high energy densities, the contributions from the charges are
dominating.

\subsection{Summary}
In this paper we analyzed the thermodynamic properties of a model of static
sources on a line
interacting through non-Abelian forces. It was shown that the partition
function takes the form of the partition function of a gauged 
principal chiral model. Using the eigenvalue density as collective field
variable the Hamiltonian for the eigenvalue density in the large $N$-limit 
was computed. We gave a
static solution of the corresponding Hamilton equations. For the special case
of only two types of charges, the static solution 
was parametrized using the parameter $\mu$ proportional to the Fermi energy. 
In particular the case of two types of 
charges transforming under the adjoint, and charges 
transforming under the fundamental representation of the gauge group was 
considered. 
Expanding the parametrized solutions at $\mu = 1$ we established the 
existence of a straight line in the phase diagram where the free 
energy exhibits a third order phase transition. We proved that the third
order behaviour terminates at a second order point. The critical line
then continues as a first order line, which was determined numerically.

The whole phase diagram was interpreted by analyzing the contributions
of charges and flux to the energy density. We found that 
for $\mu >1$ the system is characterized by low energy densities, while
for $\mu < 1$ the densities are high. The behaviour of higher winding
Polyakov loop operators provides a powerful tool for a further 
characterization of the 
high and low density phases. In the low density phase we found exponential 
suppression with increasing winding number, while for the low density 
phase we proved a power-law behaviour. 

It is tempting to analyze if the change of the behaviour of the Polyakov
loops can be seen also in deconfining phase transitions in higher dimensions,
and for finite $N$. If the picture persists, the Polyakov loop operators
would provide a  powerful criterion to analyze deconfinement phase 
transitions. In particular in lattice calculations this operator would 
be easier to analyze than other criteria as e.g. the area law of the 
Wilson loop operator.

Also the 2-dimensional model could be generalized in several directions. 
It would be interesting to analyze non-static solutions of the Hamilton
equations and different boundary conditions which might be used to 
include a $\theta$-term. Loop expansion of the fermion determinant of 
QCD$_2$ with large quark masses could be used to relate the fugacities 
of the non-Abelian gas analyzed in this article to the mass parameters
of QCD$_2$.

\subsection*{Acknowledgment}
This work was supported in part by the Natural Sciences and Engineering
Research Council of Canada. L.~P.~is supported 
in part by a University of British Columbia Graduate Fellowship. 

%
%
% APPENDIX APPENDIX APPENDIX APPENDIX APPENDIX APPENDIX
%
%
\newpage

\appendix
\renewcommand{\thesection}{Appendix}
\section{}
\renewcommand{\thesection}{A}
\setcounter{equation}{0}
\subsection{Cubic term in the Jevicki-Sakita Hamiltonian}

In this section of the appendix we bring the second term on the
left-hand-side of Equation (\ref{jsham}) to a convenient form.
The variation  by $\rho(\theta)$ of this term is given by (up to a factor)
\[
{\cal W}(\theta)=\left({\cal P}\int d\theta'\rho(\theta')
\cot\frac{ \theta-\theta'}{2}
\right)^2-2{\cal P}\int d\theta'\rho(\theta')
d\theta''\rho(\theta'')\cot\frac{\theta-\theta'}{2}
\cot\frac{\theta'-\theta''}{2} \; .
\]
In the integrals, we change variables to the complex variable
$$t=e^{i\theta} ~~,~~t'=e^{i\theta'}~~,~~t''=e^{i\theta''} \; ,$$ so that
the integrals are over an interval on the unit circle and
\[
{\cal P}\int d\theta'\rho(\theta')\cot\frac{\theta-\theta'}{2} \; = \;
{\cal P}\int \frac{dt'}{t'}\frac{ t+t'}{t-t'}\rho(t') \; = \;
i+2{\cal P}\int dt'\frac{1}{t-t'}\rho(t') \; .
\]
We obtain
\[
{\cal W}(t) \; = \; 4 \left( {\cal P}\int dt'\frac{\rho(t')}{t-t'}\right)^2
-8{\cal P}\int dt'dt''\frac{\rho(t')}{t-t'}\frac{\rho(t'')}{t'-t''}
+1-4{\cal P}\int dt'dt''\frac{\rho(t')\rho(t'')}{t'(t'-t'')} \; .
\]
The last two terms are constants (which because of the normalization
condition the the density $\rho(t)$ must satisfy, must be irrelevant),
and the first term can be found by the following argument.  We
consider the function
\[
G(z) \; = \; \int dt\frac{\rho(t)}{t-z} \; .
\]
This function is analytic everywhere in the complex plane except on
the arc of the unit circle where the eigenvalue density has support.
Obviously
\[
G(z)~\rightarrow ~ 0~~~{\rm as}~|z|\rightarrow\infty \; .
\]
Also, by letting $z$ approach the support of $\rho(t)$ from outside
and inside the unit circle, we obtain ($\varepsilon >0$)
\begin{equation}
\lim_{\varepsilon \rightarrow 0} G(t(1+\varepsilon)) \; \equiv \;
G_+(t) \; = \; {\cal P}\int dt'\frac{\rho(t')}{t'-t}~-~i\pi\rho(t) \; ,
\label{g+}
\end{equation}
and
\begin{equation}
\lim_{\varepsilon \rightarrow 0} G(t(1-\varepsilon)) \; \equiv \;
G_-(t) \; = \; {\cal P}\int dt'\frac{\rho(t')}{t'-t}~+~i\pi\rho(t) \; ,
\label{g-}
\end{equation}
respectively.
The function
\[
K(z) \; = \; G^2(z) \; - \; 2\int dt\frac{\rho(t)}{t-z}{\cal P}\int dt'
\frac{\rho(t')}{t'-t} \; ,
\]
is obviously analytic everywhere except eventually on the support of $\rho$.
Using (\ref{g+}) and (\ref{g-}) one finds that it is
continuous across the support of $\rho$ since
\[
\lim_{\varepsilon \rightarrow 0} K(t(1+\varepsilon)) \; - \;
\lim_{\varepsilon \rightarrow 0} K(t(1-\varepsilon)) ~=~0 \; .
\]
Thus $K$ is an entire function of $z$.  Furthermore, since it vanishes at
infinity, and is analytic everywhere
\[
K(z) \; = \; 0 \; .
\]
Then, taking the real part of $K$ on the support of $\rho$ gives
\[
{\cal W}(t) \; = \; 4\pi^2\rho^2(t) \; + \; {\rm const.}~~~.
\]
This is the functional derivative of the term
\begin{equation}
\frac{4\pi^2}{3}\int d\theta\rho^3(\theta)~+~{\rm const.}~~~,
\label{TFterm}
\end{equation}
which is proportional to
the second term in the Hamiltonian in the
Schr\"odinger equation (\ref{jsham2}).

\subsection{Analysis of the parametric integrals $I(\mu), M(\mu)$}

Here we analyze the properties of the parameter integrals
$I(\mu)$ and $M(\mu)$ originally defined in (\ref{Cnorm}). A transformation
of the integration variable brings the integrals to the form
\begin{eqnarray}
I(\mu) \;  & = & \; \frac{4}{\pi} \int_{0}^{\pi} \sqrt{\mu + \cos(\theta)} \;
\mbox{H}(\mu + \cos(\theta)) \; d\theta \; ,\nonumber \\
M(\mu) \; & = & \; \frac{4}{\pi} \int_{0}^{\pi} \cos{\theta}
\sqrt{\mu + \cos(\theta)} \;
\mbox{H}(\mu + \cos(\theta)) \; d\theta \; ,
\label{CAparint}
\end{eqnarray}
which shows, that they are independent of $J$. H(..) denotes the
step function. For $\mu = 1$ they can be evaluated explicitly, giving
\begin{equation}
I(1) \; = \; \frac{8 \sqrt{2}}{\pi} \; \; \; \; , \; \; \; \;
M(1) \; = \; \frac{1}{3} \frac{8 \sqrt{2}}{\pi} \; .
\label{Cintmu1}
\end{equation}
For $\mu \geq 1$, the range of integration is $[0,\pi]$, while for
$\mu < 1$, the step function reduces the range of integration to
$[0,\arccos(-\mu)]$. $I(\mu)$ can be expressed in terms of complete
$E(p)$ and incomplete $E(p;\phi)$ elliptic integrals
\begin{equation}
I(\mu) \; = \; \left\{ \matrix{
\frac{8 \sqrt{\mu + 1}}{\pi} E \left( \sqrt{\frac{2}{\mu+1}} \right) &
\mbox{for} \; \mu \geq 1 \cr
\frac{8 \sqrt{\mu + 1}}{\pi} E\left( \sqrt{\frac{2}{\mu+1}} ;
\arcsin \left( \sqrt{\frac{\mu+1}{2}}\right) \right) &
\mbox{for} \; \mu \leq 1  } \right. \; .
\label{CIell}
\end{equation}
When expanding $I(\mu)$ and $M(\mu)$
\begin{equation}
\mu = 1 \pm \varepsilon, \varepsilon > 0 \; \; \; \; ,  \; \; \; \;
I(1 \pm \varepsilon ) = I(1) + \delta_\pm I \; \; \; \; , \; \; \; \;
M(1 \pm \varepsilon ) = M(1) + \delta_\pm M \; ,
\label{CAexp}
\end{equation}
one can use for $I(\mu)$ the well known formulas for the expansion
of elliptic integrals (see e.g. \cite{Spanier}) to obtain
\begin{equation}
\frac{\delta_\pm I}{I(1)} \; \; = \; \; \mp \varepsilon \ln(\varepsilon)
\frac{1}{8} \; \; \pm  \; \; \varepsilon \frac{1}{8}[5 \ln(2) \pm 1]
\; \; + \; \; o(\varepsilon) \; .
\label{CvarI}
\end{equation}
The variation $\delta_\pm M$ can be related to $\delta_\pm I$ by using
\[
M(\mu) + \mu I(\mu) \; = \; \frac{4}{\pi} \int_0^{\pi}
[\mu + \cos{\theta} ]^{3/2} \; H( \mu + \cos(\theta) ) d\theta \; .
\]
Expanding the left hand side using (\ref{CAexp}) and the right hand
side using Taylor expansion, one obtains
\begin{equation}
\frac{\delta_\pm M}{M(1)} \; \; = \; \; -3 \frac{\delta_\pm I}{I(1)}
\; \; \pm \; \; \frac{3}{2} \varepsilon \; .
\label{CvarM}
\end{equation}
We remark, that $I(\mu)$ and $M(\mu)$ have first derivatives that
diverge logarithmically as $\mu$ approaches 1. This fact can be seen from
(\ref{CIell}), (\ref{CvarM}) and known formulas for the derivatives of
elliptic integrals. 

Finally we denote the large $\mu$ behaviour and the behaviour at $\mu \sim -1$.
In both cases the results can be computed easily by expanding (\ref{CAparint}).
We obtain
\begin{eqnarray}
I(\mu) \; &=& \; \sqrt{\mu} \; 4 \Big[ 1 - \frac{1}{16} \frac{1}{\mu^2} +
O(1/\mu^4) \Big] \; , \nonumber \\
M(\mu) \; &=& \; \frac{1}{\sqrt{\mu}} \; \Big[ 1 + O(1/\mu^2) \Big] \; ,
\label{largemu}
\end{eqnarray}
for $\mu \gg 1$, and 
\begin{eqnarray}
I(-1+\varepsilon) \; &=& \; \varepsilon \; \sqrt{2}
\; \;  + \; \; o(\varepsilon)  \; , \nonumber \\
M(-1+\varepsilon) \; &=& \; \varepsilon \; \sqrt{2}
\; \;  + \; \; o(\varepsilon) \; ,
\label{musim1}
\end{eqnarray}
for $0 < \varepsilon \ll 1$.
\subsection{Large $k$ behaviour of $M_k(\mu)$}

In this section of the 
appendix we analyze the large $k$ behaviour of the higher moments
$M_k(\mu)$ given by
\begin{equation}
M_k(\mu) \; = \; \frac{2}{\pi} \int_{-\pi}^{\pi} \cos{k \theta}
\sqrt{\mu + \cos(\theta)} \;
\mbox{H}(\mu + \cos(\theta)) \; d\theta \; .
\label{Mkmoment}
\end{equation}
Again it is straightforward to evaluate the integrals at $\mu = 1$.
We obtain
\begin{equation}
M_k(1) \; = \; \frac{8 \sqrt{2}}{\pi} (-1)^{k+1} \frac{1}{4k^2 - 1} \; .
\label{mom1}
\end{equation}
For $\mu \neq 1$ no explicit solutions are available. For the case
$\mu > 1$ we analyze the complex contour integral
\begin{equation}
I_{\cal C} \; = \; \frac{2}{ i \pi}
\int_{\cal C} z^{k-1} \sqrt{\mu + \frac{1}{2}(z + z^{-1})} dz \; .
\label{Icont}
\end{equation}
For $\mu > 1$, the argument of the square root
\[
f(z) \; \equiv \; \mu + \frac{1}{2} \Big( z + z^{-1} \Big) \; ,
\]
can become real and negative,
if and only if $z$ is real and negative, i.e. $z = -r, r \geq 0$.
In particular $f(-r)$ is zero for $r = r_\pm \equiv \mu \pm \sqrt{\mu^2 -1}$,
and $f(-r)$ is positive for $r_- < r < r_+$.

\begin{figure}
\epsfysize=3in
\epsfbox[-85 440 400 710] {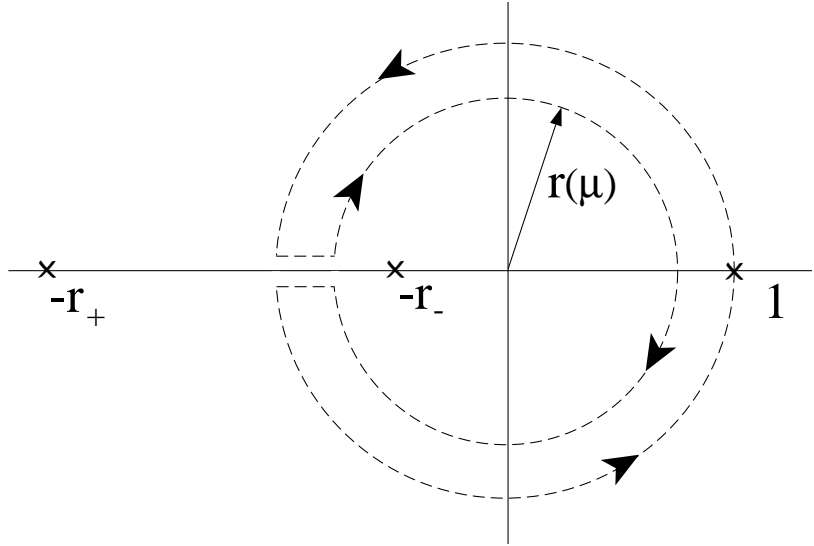}
\caption{{\it Contour for the evaluation of the integral (A.14)
\label{contplot}} }
\end{figure}

On the complex plane cut along the negative real axis
$\sqrt{f(z)}$ exists uniquely and is holomorphic.
If ${\cal C}$ is a closed curve not crossing the cut, then $I_{\cal C}$
vanishes due to Cauchy's theorem. We choose a curve as shown in
Figure \ref{contplot}. 
The outer circle is the unit circle, the inner one has radius
\[
r(\mu) \; \equiv \; \frac{1}{2} \Big( 1 + \mu - \sqrt{\mu^2 - 1} \Big) \; .
\]
The two pieces parallel to the real axis ($z = x + i \varepsilon$) cancel
in the limit $\varepsilon \rightarrow 0$ (Note that
$f(-r)$ has zeros at $r = r_\pm$.
$r(\mu)$ was chosen to lie half way between -1 and $-r_-$, so that both
integrals can be expanded in $\varepsilon$ showing that they cancel
each other.).

Thus we find that the integral along the inner circle equals the integral
along the unit circle (both integrals evaluated in counterclockwise
direction). $M_k(\mu)$ is given by the real part of $I_{\cal C}$
evaluated along the unit circle. Using the fact that the integral
along the inner circle is bounded, we establish
\begin{equation}
| M_k(\mu) | \; \leq \; \Big( r(\mu) \Big)^k \; \frac{4}{\sqrt{2}}
\sqrt{ 2 \mu + r(\mu) + r(\mu)^{-1} } \; .
\label{Aexpsup}
\end{equation}
This proves that for $\mu > 1$ the moments $M_k(\mu)$ are exponentially
suppressed with increasing $k$.

Finally we notice that for $-1 \leq \mu < 1$ exponential suppression
is excluded by well known results of Fourier transformation. Using
the fact that $M_k(\mu)$ is proportional to the Fourier coefficients
of $\sqrt{ \mu + \cos (\theta)}$ we find
\[
\sqrt{ \mu + \cos(\theta)} \; = \; \frac{1}{4} M_0(\mu)
+ \sum_{k=1}^\infty \frac{M_k(\mu)}{2} \cos(k\theta) \; .
\]
The derivative with respect to $\theta$ then is given by
\[
-\frac{1}{2} \frac{ \sin(\theta)}{\sqrt{ \mu + \cos(\theta)}} \;
= \; - \sum_{k=1}^\infty k \frac{M_k(\mu)}{2} \sin(k\theta) \; .
\]
If the $M_k(\mu)$ were exponentially suppressed, the series on the
right hand side would converge to a finite constant for all $\theta$.
This contradicts the fact, that the left hand side diverges at
$\theta = \arccos(-\mu)$ for $\mu < 1$. Thus for $-1 \leq \mu \leq 1$
(for $\mu =1$ see the result (\ref{mom1}))
we find only power-like suppression of $M_k(\mu)$ for large $k$.

\end{document}